\documentclass[12pt, onecolumn, groupedaddress, floatfix, longbibliography]{revtex4-1}

\usepackage{amsmath}
\usepackage{amssymb}
\usepackage{graphicx}
\usepackage{float}
\usepackage[english]{babel}
\usepackage{dsfont}
\usepackage{epstopdf}
\usepackage{soul}
\usepackage{color}
\usepackage{braket}
\usepackage[normalem]{ulem}
\usepackage[colorlinks=true,
            linkcolor=blue,
            urlcolor=blue,
            citecolor=blue]{hyperref}

\definecolor{SMblue}{rgb}{0,0,0}

\emergencystretch=\maxdimen
\begin{document}
\title{{Nonlinearity-induced Band Gap Transmission in Dispersive and Flat Band Photonic Lattices}}

\author{Avinash Tetarwal}
\affiliation{Department of Physics, Indian Institute of Science, Bangalore 560012, India}
\author{Shailja Sharma}
\affiliation{Department of Physics, Indian Institute of Science, Bangalore 560012, India}
\author{Sebabrata Mukherjee}
\email{mukherjee@iisc.ac.in}
\affiliation{Department of Physics, Indian Institute of Science, Bangalore 560012, India}
\date{\today}

\begin{abstract}
{\color{SMblue}
Nonlinear interactions in photonic non-dispersive (flat) bands remain largely unexplored, despite their potential to yield exotic phenomena. Here,}
we demonstrate nonlinearity-induced transport of light from a boundary waveguide into photonic lattices with dispersive and flat bands. For the one-dimensional lattice supporting a dispersive band, {\color{SMblue}self-focusing Kerr nonlinearity effectively makes the boundary waveguide phase-matched with the lattice modes,} enabling efficient energy transfer above a threshold input power. In contrast, {\color{SMblue}such nonlinear transmission to the flat band modes is inhibited, as demonstrated in a rhombic lattice supporting an isolated flat band.} Instead, as the nonlinearity increases, light couples periodically to the lattice edge mode and then gradually spreads into the lattice due to the excitation of the lower dispersive band.
\end{abstract}
\let\clearpage\relax

\maketitle

Periodic arrays of coupled optical waveguides provide a versatile platform for exploring how the transport of light is influenced by 
lattice geometry, synthetic gauge fields, disorder, and nonlinearity~\cite{christodoulides2003discretizing, Garanovich2012light, Longhi2009quantum, schwartz2007transport, segev2013anderson, Christodoulides1988}. The lattice geometry, together with synthetic gauge fields, can determine the band dispersion -- i.e., the variation of `energy' with quasi-momentum -- which influences the spreading or the diffraction of light in a lattice. On the other hand, nonlinearity can, in general, suppress diffraction, leading to the formation of shape-preserving nonlinear waves such as solitons~\cite{Christodoulides1988, segev1992spatial, eisenberg1998discrete, lederer2008discrete} and discrete breathers~\cite{flach1998discrete, kopidakis2000discrete, mandelik2003observation, shit2024probing}.  

In the context of wave transmission in a periodic medium, the presence of a band gap implies that certain states with energies within the gap can not propagate, effectively blocking the energy transmission. However, band gap transmission can be achieved by introducing nonlinearity in the system. Specifically, Ref.~\cite{F_Geniet_PRL} proposed a discrete sine-Gordon lattice driven at one end with energies within a band gap.
For low-amplitude linear driving at the edge, no energy flows through the lattice; however, above a threshold nonlinear strength, sudden energy flow occurs due to the formation of nonlinear modes. This effect, coined as nonlinear supra-transmission, is a general wave phenomenon that can be observed in various setups, including nonlinear {\color{SMblue}waveguide arrays~\cite{Ramaz, susanto2008calculated, motcheyo2017homoclinic, zakharov2023effect, susanto2023surge, MACIASDIAZ2007447, Motcheyo}.
Here, we investigate nonlinear band gap transmission using femtosecond laser-fabricated photonic lattices that are weakly coupled to a boundary waveguide with a relatively lower linear refractive index. In our experiments, intense laser pulses are coupled to the boundary waveguide, introducing self-focusing optical Kerr nonlinearity that causes an intensity-dependent increase in the refractive index. Unlike the supra-transmission models~\cite{Ramaz, susanto2008calculated, motcheyo2017homoclinic, zakharov2023effect, susanto2023surge}, the nonlinearity in our case effectively makes the boundary waveguide phase-matched with the lattice, enabling energy transfer to the lattice modes.}

{\color{SMblue}We then study nonlinear band gap transmission in a quasi-one-dimensional rhombic lattice supporting an isolated flat band. 
Compact localized flat band modes~\cite{tasaki2008hubbard, leykam2018perspective} exhibit fascinating localization effects due to their infinite effective mass.
While linear transport in flat-band photonic lattices~\cite{FB_PRL_SM_Lieb, vicencio2015observation, RhombicFB_SM, FB_Lieb_IOP, mukherjee2018experimental, xia2016demonstration, taie2015coherent, zeng2024transition, schulz2017photonic} has been studied, nonlinear effects in such systems remain largely unexplored.
Here, we experimentally and numerically show that the aforementioned nonlinear band gap transmission to the flat band modes does not occur.
However, up to a certain nonlinear strength, periodic light transfer is observed to the lattice edge mode due to the formation of discrete breathers~\cite{flach1998discrete, kopidakis2000discrete, mandelik2003observation, shit2024probing}. At higher nonlinearities, we demonstrate efficient light transfer to the lower dispersive band of the lattice.}

\begin{figure*}[]
\centering\includegraphics[width=0.8\linewidth]{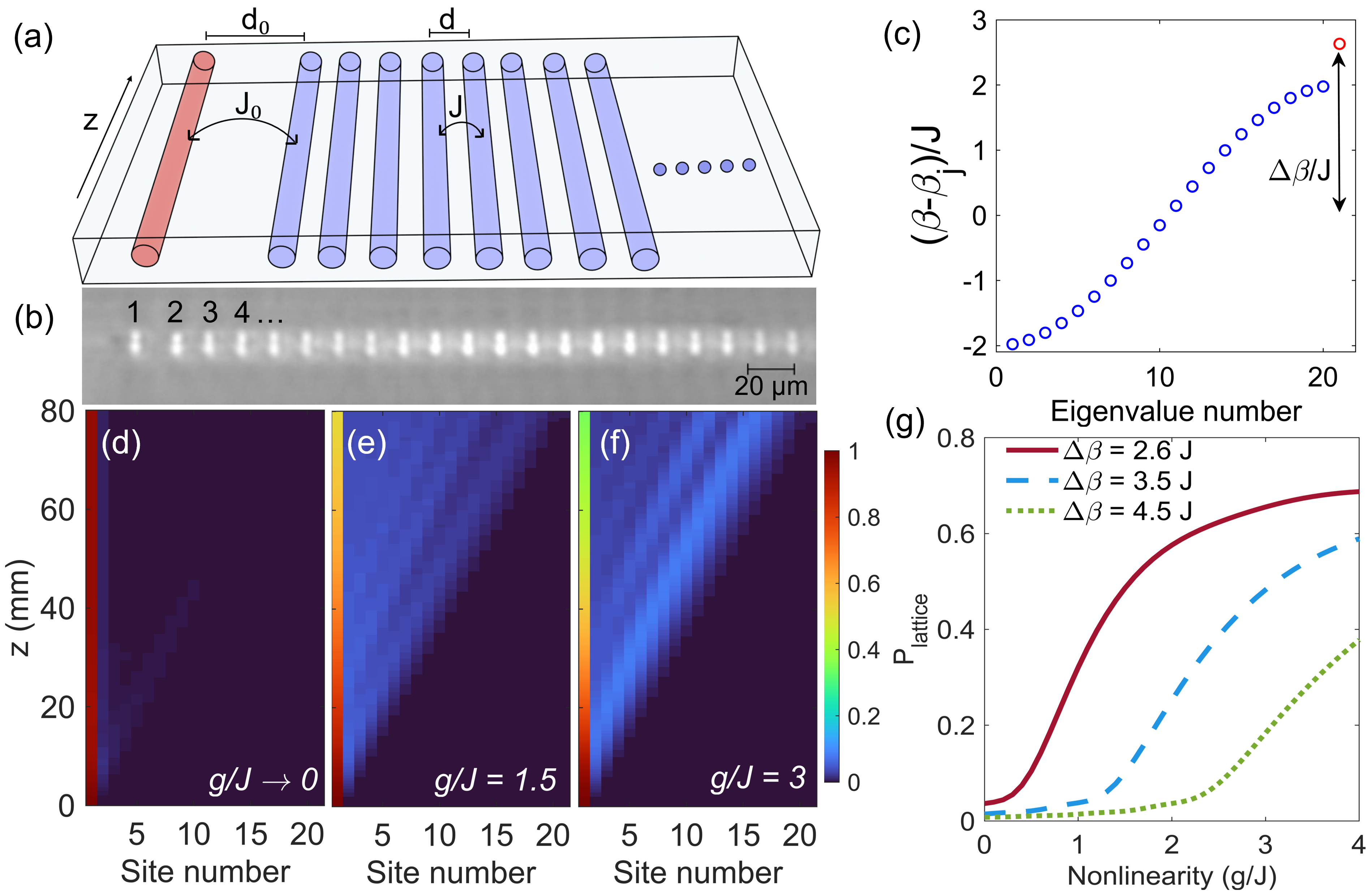}
\caption{(a) Sketch of a one-dimensional array of identical waveguides, 
each with propagation constant $\beta$ and coupling strength $J$. A boundary waveguide with a lower propagation constant $\beta_0\!<\!\beta$ is weakly connected to the array with a smaller coupling $J_0\!<\!J$.
(b) Micrograph (cross-section) of the photonic device fabricated using femtosecond laser-writing.
(c) Numerically calculated eigenvalue spectrum of the device, consisting of $20$ sites in the lattice. {\color{SMblue}The red circle corresponds to the boundary mode.} Here, $\Delta \beta\!\equiv\!\beta-\beta_0\!=\!2.6J$, and $J_0\!=\!0.29J$. %
(d-f) Calculated light propagation through the device  
for three different nonlinear strengths $g/J$, with input excitation
at the boundary waveguide. 
(g) Variation of power in the lattice $P_{\text{lattice}}$ as a function of $g/J$ for three different $\Delta \beta$. Nonlinearity causes light transmission to the lattice above a threshold nonlinearity of $g_{\text{th}}\!=\!\Delta\beta-2J$ in each case.
}
\label{Figure1}
\end{figure*}

In the scalar-paraxial approximation, light propagation through evanescently coupled waveguide networks can be described by the following discrete nonlinear Schr{\"o}dinger equation~\cite{Longhi2009quantum, Garanovich2012light}:
\begin{equation}\label{eq:equation1}
i\frac{\partial}{\partial z}\psi_s(z) =  \sum_{s'} H_{ss'}^{\text{lin}}\psi_{s'}-g|\psi_s|^2 \psi_s\, ,
\end{equation}
where $z$ is the propagation distance, $s$ labels the waveguides, $H_{ss'}^{\text{lin}}$ 
are the elements of the linear tight-binding Hamiltonian, and $\psi_s$ is proportional to the slowly varying complex amplitude of the optical field at the $s$-{th} waveguide. 
The nonlinear parameter $g\!=\!2\pi n_2/(\lambda A_{\text{eff}})$ is determined by the nonlinear refractive index coefficient $n_2$, the effective area of the waveguide modes $A_{\text{eff}}$, and the wavelength $\lambda$ of light~\cite{agrawal2000nonlinear}. In the absence of optical losses, the total energy and the renormalized power ($P=\sum_s|\psi_s|^2$) are conserved. Note that the nonlinearity in the off-diagonal coupling terms is negligible in our experiments.

We first consider a one-dimensional photonic
lattice consisting of $N$ identical single-mode waveguides with nearest-neighbor coupling $J$. One end of the lattice is weakly coupled to a boundary waveguide, with coupling strength $J_0\!<\!J$.
This boundary waveguide has a 
{\color{SMblue}lower propagation constant $\beta_0$ compared to the waveguides in the lattice, such that $\Delta\beta\!\equiv \beta-\beta_0 \!>2J$.}
In Fig.~\ref{Figure1}(a),
the lattice and the boundary sites are highlighted in blue and red, respectively. 
Fig.~\ref{Figure1}(c) shows the spectrum of the device for the experimentally realized parameters, $J\!=\!0.12$~mm$^{-1}$, $J_0\!=\!0.29J$, and $\Delta \beta=2.6J$. {\color{SMblue}The propagation constant of the linear lattice modes $\beta_j$ (where $j\!=\!1, 2, \dots, N+1$)  spans from $\beta+2J$ to $\beta-2J$,} while the boundary mode (red) lies in the upper band gap.  
{\color{SMblue}Note that a small $J_0$ value ensures minimal impact of the boundary waveguide on the lattice spectrum.}
Figs.~\ref {Figure1}(d-f) show numerically calculated linear and nonlinear light propagation through the lattice over a propagation length of $80$~mm. In the linear regime 
$(g\rightarrow0)$, the initial state coupled into the boundary waveguide remains mostly localized
as shown in Fig.~\ref{Figure1}(d).
Indeed, light transport from the boundary waveguide to the lattice is suppressed in the linear regime when $\Delta\beta>2J$.
However, the introduction of nonlinearity can enable transmission to the lattice, as shown in Figs.~\ref{Figure1}(e, f), for $g/J\!=\!1.5$ and $3$.
Fig.~\ref{Figure1}(g) illustrates how the power transmitted to the lattice varies with the nonlinearity strength $g/J$ for three different values of $\Delta \beta\!=\!2.6J, \, 3.5J$ and $4.5J$. Notice the sharp increase in transmission above a threshold nonlinearity of $\Delta\beta-2J$ in each case. It should be highlighted that similar nonlinearity-induced band gap transmission above a threshold nonlinearity occurs in supra-transmission models~\cite{Ramaz, susanto2008calculated, motcheyo2017homoclinic, zakharov2023effect, susanto2023surge}, where the boundary waveguide is essentially linear.

\begin{figure*}[]
\centering\includegraphics[width=0.8\linewidth]{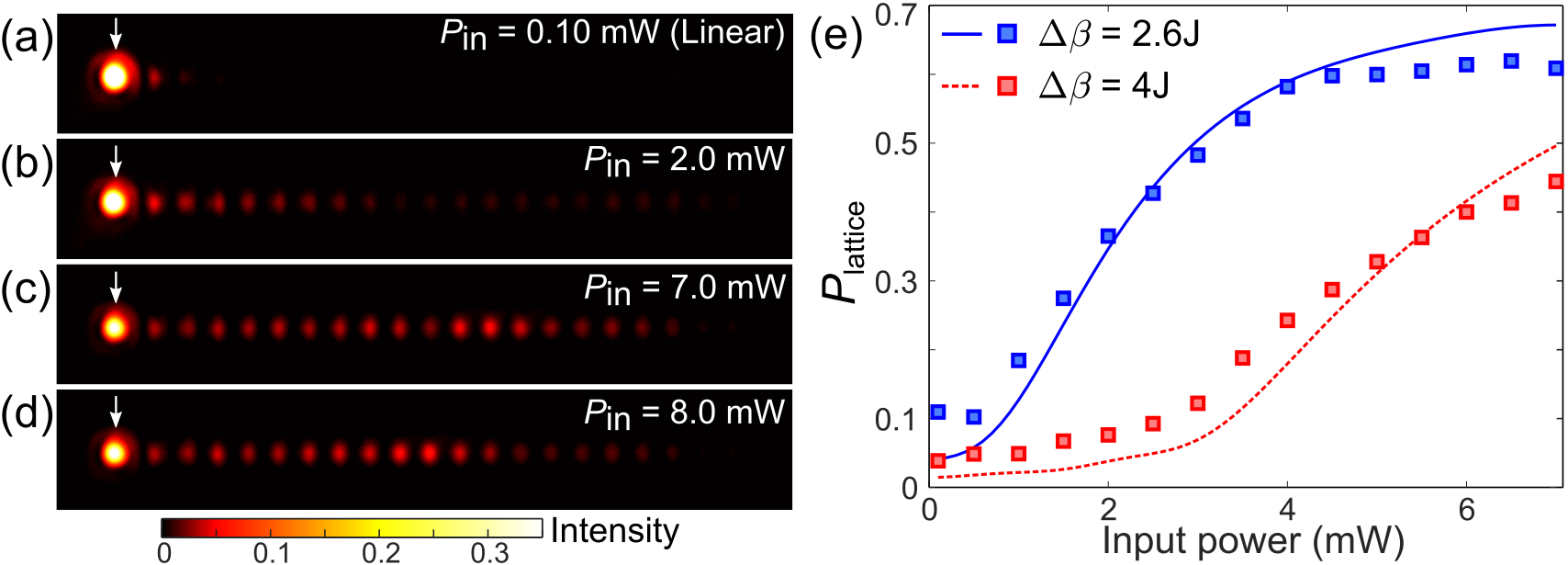}
\caption{(a-d) Experimentally measured output intensity distributions at $z\!=\!76.2$ mm for four different input powers, as indicated on each image. 
The white arrow marks the waveguide where light is initially launched.
Above a threshold nonlinearity, light transmission in the lattice increases.
(e) The measured variation of the normalized power 
$P_{\text{lattice}}$ in the lattice as a function of input power for two different values of $\Delta \beta/J\!=\! \{2.6, \, 4\}$.
{\color{SMblue} The filled squares (lines) are obtained experimentally (numerically).}}
\label{Figure2}
\end{figure*}

To demonstrate the above-mentioned nonlinear band gap transmission, we fabricate optical devices in a $76.2$~mm-long borosilicate glass (Corning Eagle-XG) substrate using the well-established technique of femtosecond laser-writing~\cite{Ams, Davis}. The
substrate is mounted on
Aerotech $x$-$y$-$z$ translation stages, and each waveguide %
is created
by translating the substrate once through the focus of a $500$ kHz train of %
$260$ fs laser pulses generated by a fiber laser system (Satsuma, Amplitude).
The fabrication parameters are optimized to inscribe well-confined, single-mode waveguides at an operational wavelength of $1030$~nm. The evanescent coupling in the device is determined by the inter-waveguide spacing, and the relative propagation constant of the boundary waveguide is controlled by adjusting the translation speed of fabrication.

We fabricate $13$ sets of photonic devices, as shown in Fig.~\ref{Figure1}(a, b), each with $d\!=\!16.5 \, \mu$m and $d_0\!=\!22 \, \mu$m, where $\{d, \, d_0\}$ are the waveguide spacing in the lattice and the spacing from the boundary waveguide to the lattice, respectively. 
All $N\!=\!20$ waveguides in each lattice are inscribed at a translation speed of $6$~mm/s, while the fabrication speed for the boundary waveguide is varied from $6$ to $12$~mm/s in steps of $0.5$~mm/s.  
{\color{SMblue}As discussed in the Supplemental document,
the propagation constant detuning is calibrated as} $\Delta\beta\!=\!0.085\Delta v$, where $\Delta v$ is the difference in translation speed for fabricating the lattice and the boundary waveguide.
For nonlinear experiments, we use the devices with the boundary site fabricated at $\{9.5, \, 11.5\}$~mm/s for which the value of $\Delta \beta$ is estimated to be $2.6J$ and $4J$, respectively.

\begin{figure*}[]
\centering\includegraphics[width=0.7\linewidth]{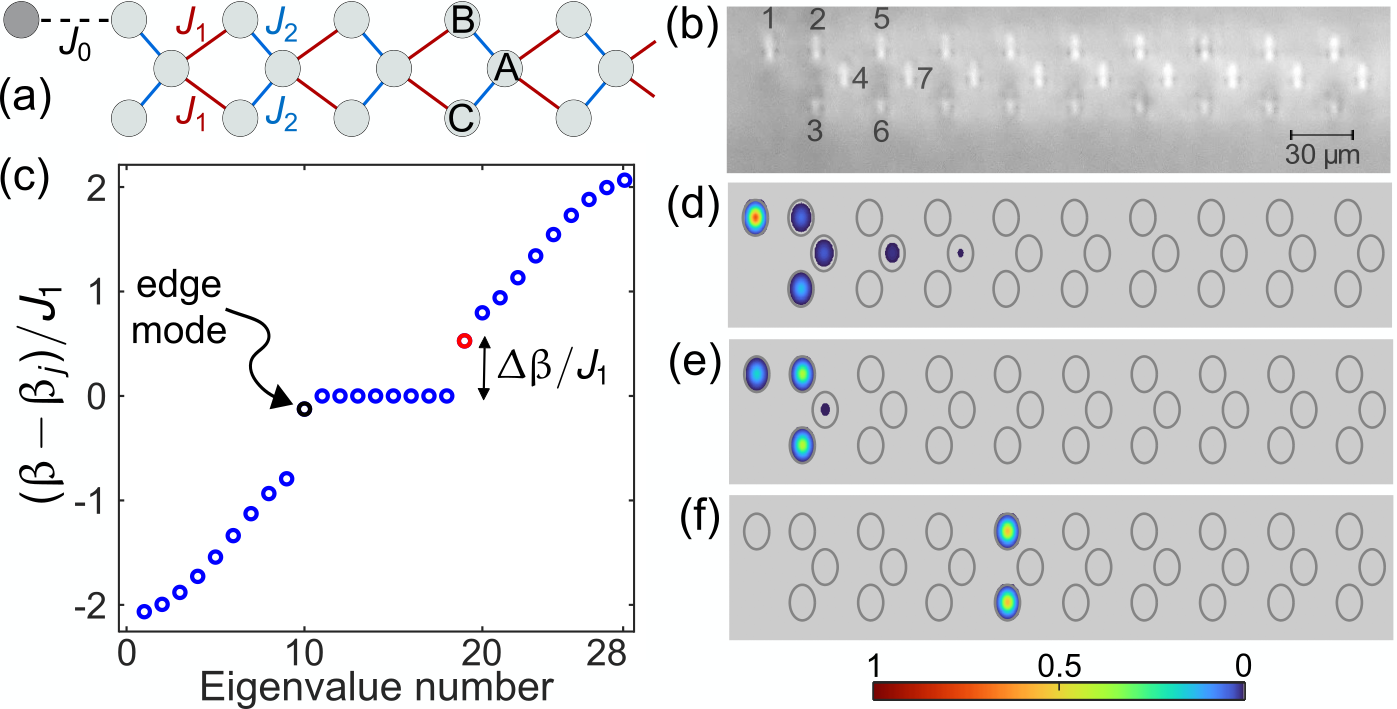}
\caption{(a) Sketch of the photonic flat band rhombic lattice weakly coupled to a boundary waveguide with a propagation constant $\beta_0\!<\!\beta$. %
(b) Transmission micrograph of the laser-fabricated photonic rhombic lattice. (c) Eigenvalue spectrum of the device, consisting of $28$ sites. The lattice supports three bands -- the upper and lower bands are dispersive, and the middle band is perfectly flat. Here, $\beta_0$ is adjusted such that the boundary mode (red) is detuned from the flat-band modes by $\Delta$.
{\color{SMblue}Intensity distributions of the (d) boundary mode, (e) lattice edge mode and, (f) compact localized flat band mode.}
}
\label{Figure3}
\end{figure*}

\begin{figure*}[t!]
\centering\includegraphics[width=0.75\linewidth]{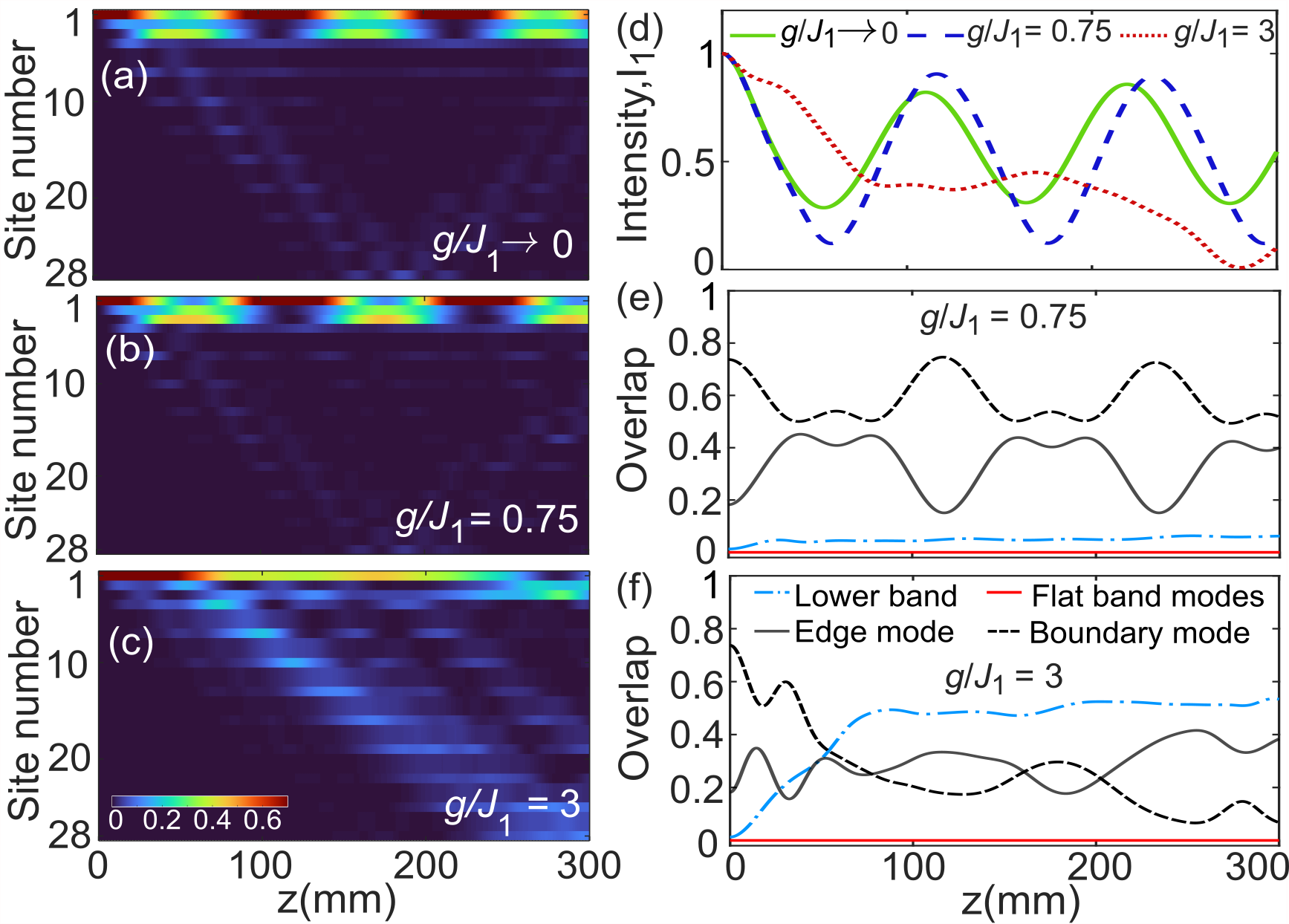}
\caption{(a-c) Numerically calculated light propagation through the rhombic lattice for three different nonlinear strengths $g/J_1$, as indicated on each image. (d) Variation of intensity in the boundary waveguide as a function of propagation distance. {\color{SMblue}(e, f) Overlap of the normalized optical state with different eigenmodes of the system for two different $g/J_1$ values.}
}
\label{Figure4}
\end{figure*}

To access self-focusing Kerr nonlinearity, we use temporally stretched, down-chirped $1.1$~ps laser pulses with tunable pulse energy. For all transport experiments, we use horizontally polarized light at 1030~nm wavelength.
Fig.~\ref{Figure2}(a) shows the measured output intensity distribution trapped at the boundary site in the linear regime. In this case,
power transfer to the lattice is minimal -- meaning, the boundary waveguide is effectively decoupled from the lattice due to its large propagation constant offset. As the nonlinearity is increased, we observe an increase in the total power in the lattice %
{\color{SMblue}$P_{\text{lattice}}\!=\!\sum_{s=2}^{21}|\psi_s|^2/ \sum_{s=1}^{21}|\psi_s|^2$}, as shown in Figs.~\ref{Figure2}(b-d). The variation of $P_{\text{lattice}}$ 
with the average input power is shown in Fig.~\ref{Figure2}(e) (red data set). 
We perform similar experiments with a larger $\Delta \beta\!=\!4J$ and observe nonlinear band gap transmission with a relatively larger threshold power, as would be expected; see the blue data set in Fig.~\ref{Figure2}(e). {\color{SMblue}The solid and dashed lines in this figure were obtained numerically (see Supplemental document).}
We note that these devices can be used as ultra-fast all-optical nonlinear switches~\cite{friberg1987ultrafast, demetriou2017nonlinear, rajeevan2025nonlinear} where the threshold input power can be adjusted by varying $\Delta\beta$. {\color{SMblue}We also note that the maximum power transfer to the lattice can be enhanced beyond $95\%$ by adjusting the system parameters such as the couplings, $\Delta\beta$, and propagation distance.
}

We now consider a quasi-one-dimensional rhombic array with three sites A, B, and C per unit cell. The inter- and intra-cell %
couplings are denoted by $J_1$ and $J_2$, as shown in Fig.~\ref{Figure3}(a). In the nearest-neighbor tight-binding approximation, the lattice supports a perfectly flat band 
and two dispersive bands with eigenvalues 
\begin{eqnarray}
\label{Eq_FB}
\epsilon_0(k)&=&0 \, , \quad \,   
{\text{and}} \quad \, \nonumber \\
\epsilon_{\pm}(k)&=&\pm \sqrt{2(J_1^2 + J_2^2 + 2J_1 J_2 \cos(ka))}\, ,
\end{eqnarray}
respectively, where $k$ is the quasi-momentum and $a$ is the lattice constant. 
Unlike in previous experiments~\cite{RhombicFB_SM, mukherjee2018experimental},
the bipartite nature of $J_{1}$ and $J_{2}$ makes the flat band isolated from the upper and lower dispersive bands.
In our experiments, the lattice consists of $N\!=\!27$ waveguides with $J_1\!=\!0.088$~mm$^{-1}$ and $J_2\!=\!0.042$~mm$^{-1}$. A boundary waveguide is weakly coupled to the B site at the end of the lattice, Fig.~\ref{Figure3}(a, b). The propagation constant difference of the boundary waveguide and its weak coupling to the lattice are $\Delta \beta\!=\!0.42J_1$ and $J_0\!=\!0.37J_1$, respectively. 
Fig.~\ref{Figure3}(c) shows the real-space spectrum of the device calculated using the experimentally realized parameters. 
{\color{SMblue}The intensity distributions of the boundary mode, lattice edge mode, and a compact localized flat band mode are shown in Fig.~~\ref{Figure3}(d-f), respectively. The flat band modes are localized to the B and C sites of a unit cell with equal intensity and opposite phases. 
}
The boundary mode spectrally resides in the gap above the flat {\color{SMblue}band, whereas the} lattice edge mode 
appears just below the flat band.
{\color{SMblue}Note that the boundary mode does not spatially overlap with
the localized flat band modes.}

To explore nonlinearity-induced transport from the boundary waveguide to the flat band lattice, we numerically calculate light propagation %
{\color{SMblue}for three different $g/J_1$ values}. In the linear regime, an input state localized at the boundary waveguide overlaps primarily with the lattice edge mode. As a result, a beating motion of intensity is observed with a partial transfer of light from the boundary to the B and C sites on the edge; see Figs.~\ref{Figure4}(a, d). As the nonlinear strength is increased, a similar beating motion of intensity is observed with a longer period and increased energy transfer to the lattice edge mode; see Figs.~\ref{Figure4}(b, d). 
Specifically, for $g/J_1 \lesssim 1.2$, the $z$-periodic oscillation of optical intensity occurs due to the formation of discrete breathers on the edge of the lattice. We note that the discrete breathers are $z$-periodic localized nonlinear states which are distinct from shape-preserving spatial solitons.
When the nonlinearity is further increased, modes of the lower dispersive band are excited; hence, we observe light transport to the lattice; Figs.~\ref{Figure4}(c, d). Evidently, nonlinearity causes efficient transfer of light to the lattice edge mode and to the lower dispersive band, but the flat band modes of the lattice are not excited. 
{\color{SMblue}
To further investigate, we
calculate the overlap of the normalized optical state with different
eigenmodes of the flat band system for two different $g/J_1\!=\!0.75$ and $3$, as presented in Figs.~\ref{Figure4}(e, f).
At low nonlinearity, the oscillating nature of the overlap with the lattice edge mode and the boundary mode indicates the periodic exchange of optical power between them. This behaviour is destroyed in Fig.~\ref{Figure4}(f), where the overlap with the lower band is significant.
Importantly, notice that the total overlap with the flat band modes (solid red line) remains zero irrespective of the $g/J_1$ value.}
The light transfer to the flat band modes is inhibited because of the unique phase and intensity profiles of the flat band modes, which make their overlap with the boundary mode zero. In other words, the coupling between the boundary mode and the flat band modes remains zero irrespective of their propagation constant mismatch.

\begin{figure*}[]
\centering\includegraphics[width=0.75\linewidth]{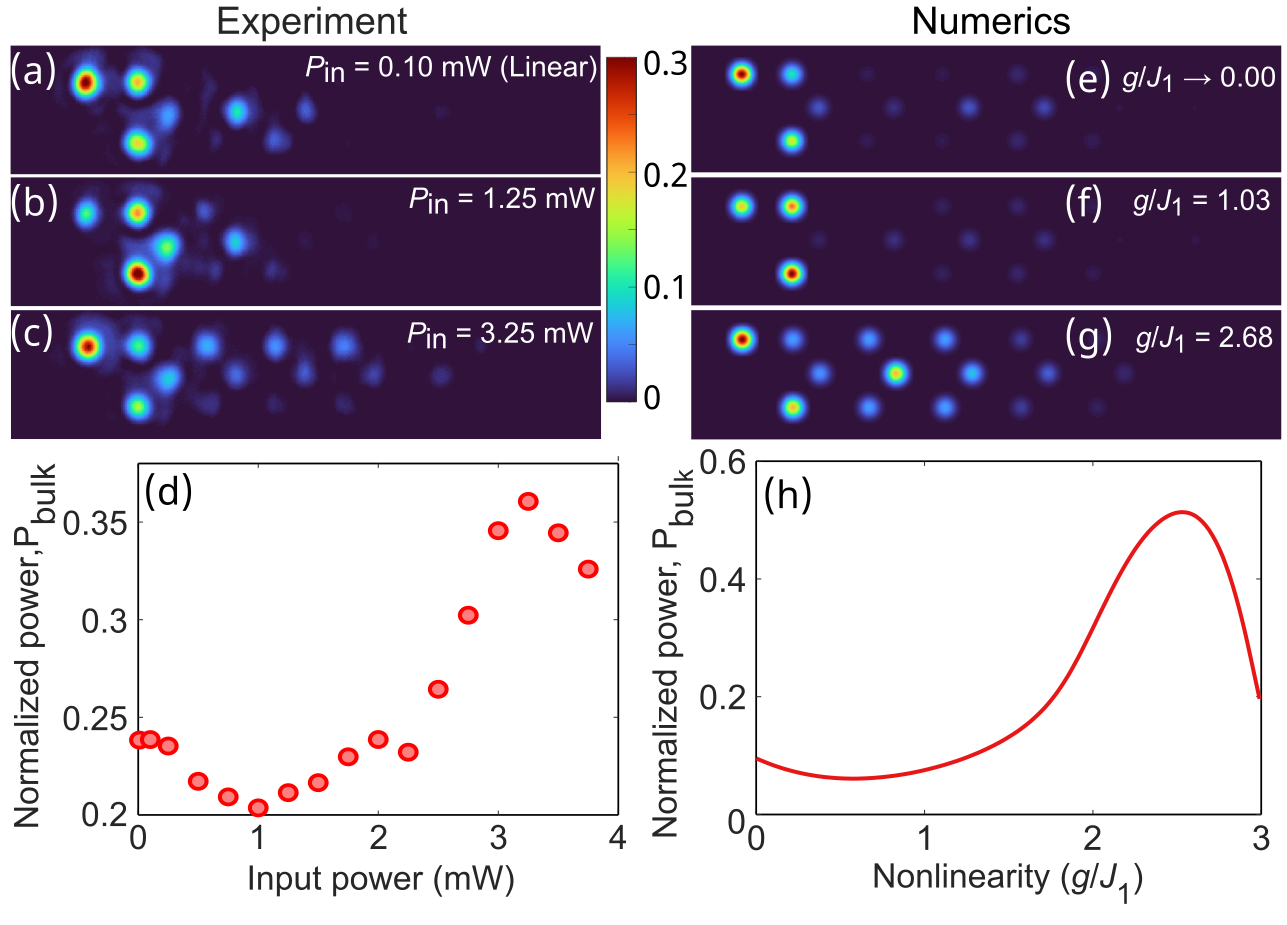}
\caption{(a-c) Experimentally measured normalized intensity at $z \!=\! 76.2$ mm for three different values of $P_{\text{in}}$. The input power is indicated on each image. 
(e-g) Numerically calculated output intensity distributions associated with (a-c).
(d, h) Measured and calculated variation of the power transmitted into the bulk of the lattice $P_{\text{bulk}}$ as a function of the nonlinear strength. 
}
\label{Figure5}
\end{figure*}

To demonstrate the signature of the nonlinear phenomena in Fig.~\ref{Figure4}, we launch laser pulses at the boundary site of the rhombic lattice and probe the intensity patterns at $z\!=\!76.2$~mm.
Figs.~\ref{Figure5}(a-c) show the measured intensity distributions 
for three different average input powers, as indicated on each figure.
In the linear regime, most of the light stays in the boundary waveguide and in the edge sites of the lattice; see Fig.~\ref{Figure5}(a). %
Up to $P_{\text{in}}\!=\!1.5$ mW, no significant spreading of light into the lattice is observed. However, notice the increased transfer of light from the boundary to the edge waveguides in Fig.~\ref{Figure5}(b), indicating the formation of nonlinear breathers.
On the other hand, the spreading of light into the lattice is clearly visible in Fig.~\ref{Figure5}(c) -- here, nonlinearity causes the excitation of modes of the lower dispersive band.  
To further clarify, we calculate and measure the light in the bulk of the lattice $P_{\text{bulk}}\!=\!\sum_{s=5}^{28}|\psi_s|^2/ \sum_{s=1}^{28}|\psi_s|^2$. Notice that $P_{\text{bulk}}$ increases only after a threshold nonlinearity, when the lower dispersive band is excited, Fig.~\ref{Figure5}(d, h). {\color{SMblue}
At strong nonlinearities ($g/J_1 \gtrsim 2.5$), the nonlinear refractive index detuning at the boundary waveguide becomes large enough to suppress power transmission through the lattice, leading to a peak in $P_{\text{bulk}}$ as a function of $g/J_1$.
We notice some deviation and asymmetry in our experiments
due to unavoidable random disorder and possible fabrication errors. That said,} Fig.~\ref{Figure5}(a-d) qualitatively agrees with the numerical results in Fig.~\ref{Figure5}(e-h).


In summary, we have numerically and experimentally studied nonlinearity-induced band gap transmission from a boundary waveguide to photonic lattices. 
When the linear propagation constant of the boundary waveguide is in the band gap, light transfer to a one-dimensional lattice is possible above a threshold nonlinear strength. 
This work can find future applications in ultra-fast all-optical nonlinear switching and signal processing.
In the case of a flat band rhombic lattice, we show that the nonlinearity can not couple light from the boundary waveguide to the flat band modes. We have observed periodic light transfer to the edge mode, and, subsequently, to the lower dispersive band of the lattice. 
For weak coupling $J_0$ to the lattice and short propagation distances (i.e., when the power depletion at the boundary site is not significant), optical nonlinearity can be used to selectively excite modes of a dispersive band.
{\color{SMblue}Evidently, photonic lattices provide a natural platform for exploring nonlinear interactions in flat bands, where such interactions can be enhanced, opening new avenues for research~\cite{cao2018unconventional, di2019nonlinear, gligoric2019nonlinear, goblot2019nonlinear, pelegri2020interaction}. Furthermore,} our results will be useful to the fundamental science of discrete solitons and breathers. \\

\noindent {\it Funding.$-$} STARS, Ministry of Education, Government of India (MoESTARS/STARS-2/2023-0716); Indian Institute of Science (Start-up grant, IoE postdoctoral fellowship); Infosys Foundation, Bangalore; Council of Scientific \& Industrial Research (PhD Scholarship). \\ \vspace{-0.6cm}

\noindent {\it Acknowledgments.$-$}
We sincerely thank Nicholas Smith of Corning Inc.~for providing high-quality glass wafers used in the experiments.\\ \vspace{-0.6cm}

\noindent {\it Disclosures.$-$} The authors declare no conflicts of interest.\\ \vspace{-0.6cm}

\noindent {\it Data Availability.$-$} The data that support the findings of this study are available from the corresponding author upon reasonable request.\\



        \counterwithout{equation}{section}
        \renewcommand{\theequation}{A\arabic{equation}}%
        \setcounter{equation}{0}
         \setcounter{section}{0}
         \renewcommand{\thesection}{\Alph{section}}%

\section*{{Supplementary Document}}

\section{Band structure of the Flat band lattice}

Light transport in our photonic lattices can be approximated by the coupled-mode equations. In this case, the Fourier- transformed Hamiltonian
in $k$-space is expressed as the following $3 \times 3$ matrix 
\begin{equation} \label{S_H}
\begin{aligned}
{\mathcal{H}} = -\begin{bmatrix}
\beta_a & (J_1 + J_2e^{ika}) & (J_1 + J_2e^{ika}) \\
(J_1 + J_2e^{-ika}) & \beta_b & 0 \\
(J_1 + J_2e^{-ika}) & 0 & \beta_c
\end{bmatrix}
\end{aligned} \, 
\end{equation}
where $J_{1, 2}$ are the inter-waveguide couplings described in the main text and $\beta_{a-c}$ are the propagation constants of {A, B,} and {C} sites of the unit cell. Since, all the waveguides in the rhombic lattice (not the boundary site) are fabricated using identical parameters, we consider {\color{SMblue}$\beta_a \!=\! \beta_b \!=\! \beta_c\!=\!\beta$.} 
In this situation, by diagonalizing the Fourier-transformed Hamiltonian, we obtain the 
{\color{SMblue}eigenvalues as $0$ and $\pm \sqrt{2(J_1^2 + J_2^2 + 2J_1 J_2 \cos(ka))}$ for the flat band and two dispersive bands, respectively.}

\begin{figure}[h!]
\centering
\includegraphics[width=0.5\textwidth]{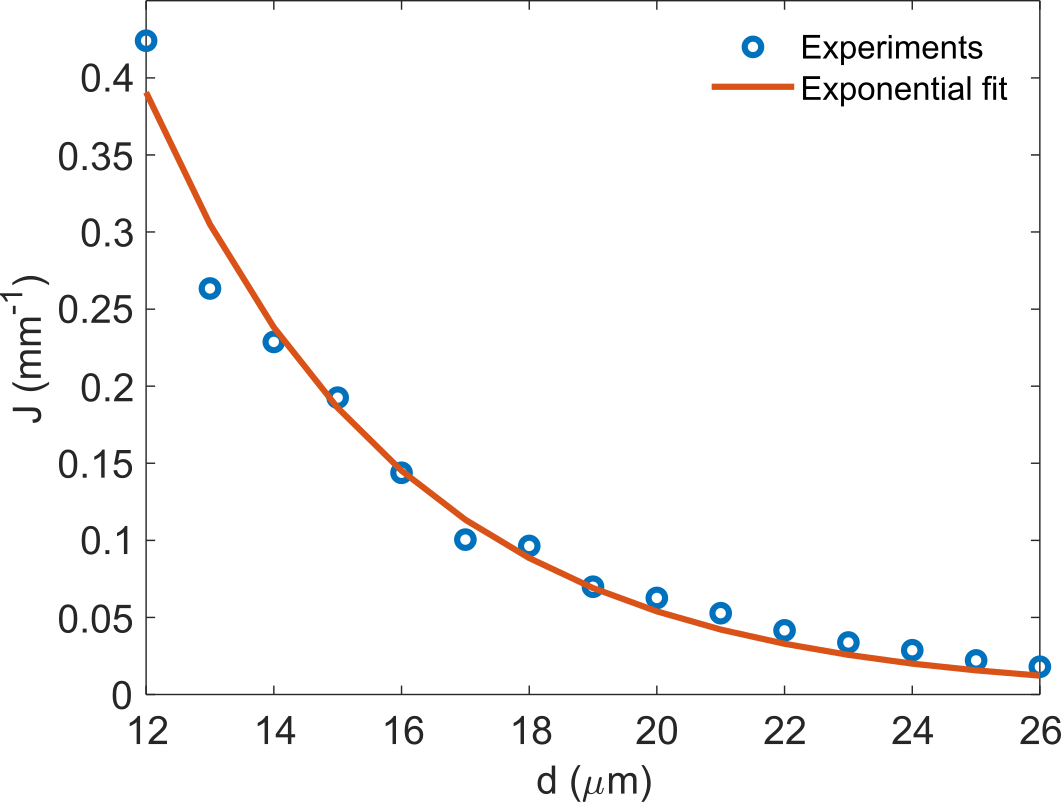}
\caption{{\color{SMblue}Variation of coupling strengths at $1030$~nm wavelength with waveguide-to-waveguide separation $d$. 
}}
\label{Figure_8}
\end{figure}

\section{Measurement of coupling strength} 
{\color{SMblue}To investigate the variation of coupling strength $J$ with the inter-waveguide separation $d$, we fabricated fifteen sets of directional couplers consisting of horizontally-coupled two straight waveguides. The separation $d$ was systematically varied from 12~$\mu$m to 26~$\mu$m in steps of 1~$\mu$m. %
The measured variation of $J(d)$ is presented in  Fig.~\ref{Figure_8}. %
Evidently, the coupling strength decays exponentially with inter-waveguide spacing, as would be expected. Using  Fig.~\ref{Figure_8} as a reference, we fabricated three sets of identical couplers with specific $d$-values, and the average coupling strengths are mentioned in the main text. We performed similar experiments to obtain couplings along the diagonal direction in the case of rhombic lattices.
}

\begin{figure}[]
\centering\includegraphics[width=0.85\linewidth]{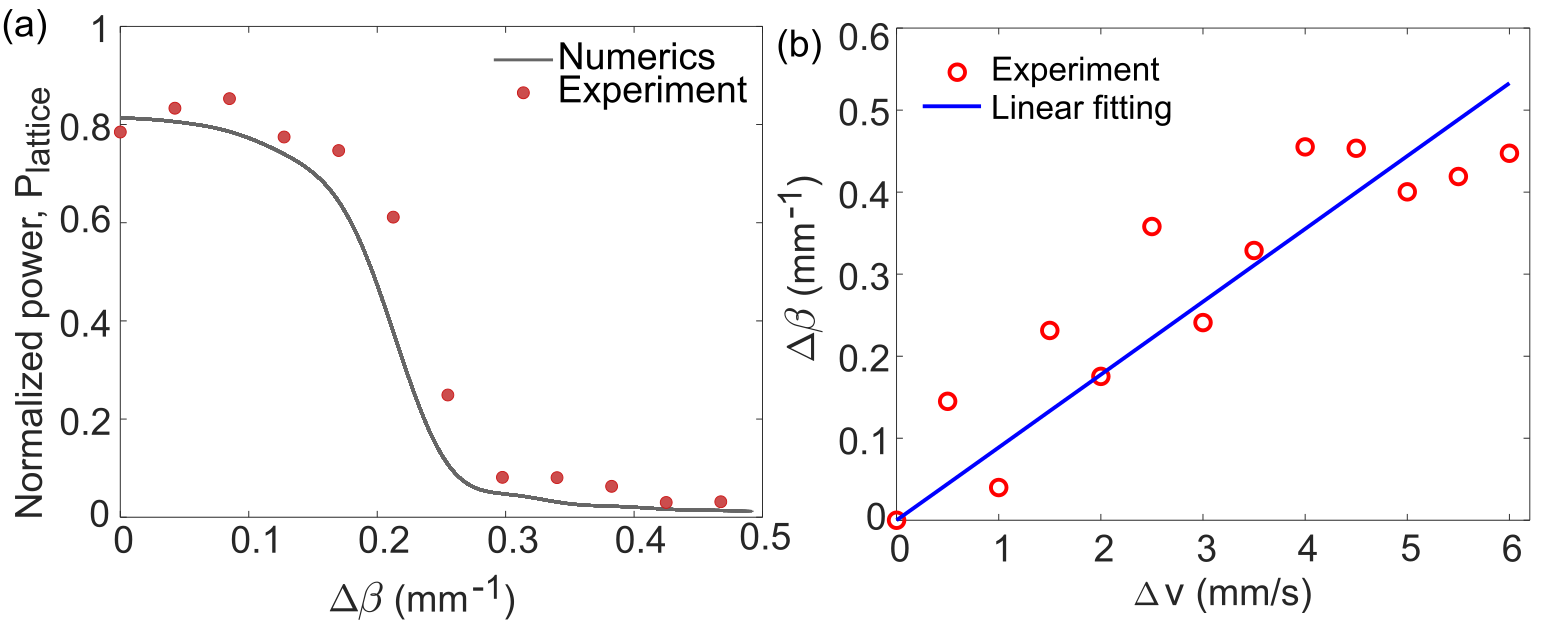}
\caption{(a) The normalized intensity in the lattice as a function of $\Delta\beta$, which is experimentally tuned by varying the translation speed. (b) The variation of $\Delta\beta$ as a function of the change in fabrication speed. The red circles are the measured values of $\Delta\beta$, and the solid blue line is a linear fit. %
 }
\label{Figure6}
\end{figure}

\section{Calibration of $\Delta\beta$}
As described in the main text, we fabricated $13$ sets of photonic devices, as shown in Fig.~1(a, b),
varying the propagation constant shift $\Delta\beta$ of the boundary site. In experiments, $\Delta\beta\!=\!\beta-\beta_0$ is increased by fabricating the boundary waveguide at a higher translation speed. 
{\color{SMblue}We couple low-power light at the boundary waveguide of each device and measure the output intensity distributions at $z\!=\!76.2$~mm. For the device with $\Delta \beta \!=\!0$, around $80\%$ of light transmits to the lattice. }
As the propagation constant of the boundary waveguide is decreased by increasing its fabrication speed, power transmission in the lattice $P_{\text{lattice}}$ %
decreases. The variation of $P_{\text{lattice}}$ with translation speed is shown in Fig.~\ref{Figure6}(a). 
From this variation of $P_{\text{lattice}}$ with translation speed, we calibrate the propagation constant detuning as $\Delta\beta\!=\!0.085\Delta v$, where $\Delta v$ is the difference in translation speed for fabricating the lattice  and the boundary waveguide. The solid line in Fig.~\ref{Figure6} (a) was calculated numerically by solving the coupled-mode Eq.1
in the linear regime ($g \! \rightarrow \!0$).

In addition to the above approach, we calibrate $\Delta \beta$ by characterizing asymmetric directional couplers.
To measure $\Delta\beta$ with translation speed, thirteen sets of isolated straight
waveguide couplers (translation speeds 6 to 12 mm/s in steps of 0.5 mm/s) were fabricated. We injected low-power light beam in one of the waveguides (WG-1) of the coupler and measured the output intensity distribution. 

We note that the output intensity distribution in an asymmetric coupler is given by
\begin{equation} 
|\psi_1(z)|^2 = 1 - 
\frac{J^2}{
    \frac{1}{4}(\Delta\beta)^2 + J^2
}
\sin^2 \left[
    \left\{
        \left( \frac{1}{4}(\Delta\beta)^2 + J^2 \right)^{\frac{1}{2}} z
    \right\}
\right]
\end{equation} 
\begin{equation} 
|\psi_2(z)|^2 = 1 - |\psi_1(z)|^2
\end{equation} 
For our laser-written couplers with known coupling strength $J$, propagation distance $z$ and output intensities $|\psi_{1,2}(z)|^{2}$, we can estimate $\Delta\beta$ as a function of $\Delta v$, see Fig.~\ref{Figure6}(b). This second approach gives a very similar calibration of propagation constant shift, $\Delta\beta\!=\!0.088\Delta v$.

\section{Loss measurement} 
The propagation loss of the laser-written waveguides is measured using the cut-back method. The insertion loss (arising due to propagation, coupling and Fresnel reflection losses) is measured for waveguides with two different lengths, i.e., $76.2$~mm and $20$~mm. From the difference in the insertion losses for the waveguides with these two different lengths, we obtain propagation loss coefficient, $\alpha \approx 0.04\ \mathrm{cm}^{-1}$.

For all nonlinear experiments, we monitor both input and output power. The output power is found to vary linearly with the input power in our experiments. This observation indicates that the nonlinear loss due to multi-photon absorption can be neglected.

\section{Calibration of nonlinear strength}
To estimate how the nonlinear strength at the input, $g(z\!=\!0)$, varies as a function of average input power $P_{\text{in}}$, we probe the nonlinear dynamics in devices shown in Fig.~1(a). 
The light beam is coupled into the boundary waveguide of the device, and the normalized power in the lattice $P_{\text{lattice}}(z\!=\! 76.2\, {\text{mm}})$ is measured as a function of the input power. The experimentally measured data for $\Delta \beta/J \!=\! \{2.6, \, 4 \}$ %
{\color{SMblue}in Figs.~2(e)}
are fitted with numerically calculated results by varying $g(0)$ as a free parameter. For both devices with two different values of $\Delta$, we obtain $ g(z\!=\!0) \!=\! 0.072 \, P_{\text{in}}$~mm$^{-1}$~mW$^{-1}$.

\begin{figure*}[t!]
\centering\includegraphics[width=1\linewidth]{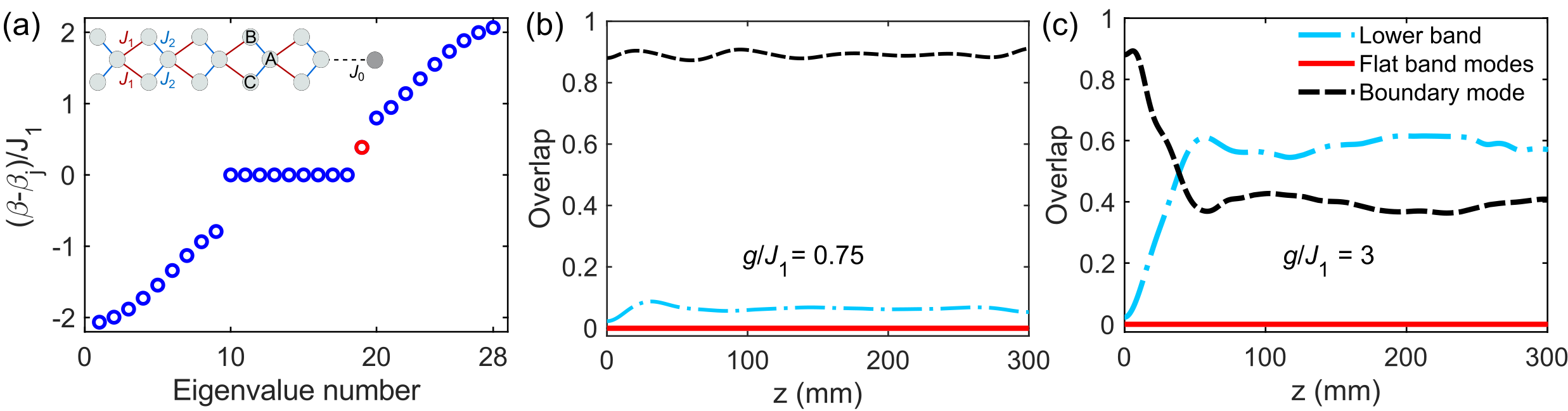}
\caption{{\color{SMblue}(a) Numerically calculated eigenvalue spectrum
of the device, consisting of 28 sites. The boundary waveguide is weakly coupled to the edge A site
of the rhombic lattice. The inset shows a sketch of the system. Note that the boundary mode lies in the band gap above the flat band, and there is no lattice edge mode in the band gap. (b, c)  Overlap of the normalized optical state with different eigenmodes of the system for two different $g/J1$ values indicated in each figure.}
 }
\label{Figure_Asite}
\end{figure*}

{\color{SMblue}
\section{Boundary waveguide Coupled to the edge A site}

In the main text, we have numerically and experimentally studied the nonlinearity-induced transport of light from a boundary waveguide to the rhombic lattice.
In that case, the boundary waveguide was coupled to the edge B site simply because the flat band modes are localized on the B and C sites of a unit cell. In this section, we numerically examine the case where the boundary waveguide is coupled to the edge A site as shown in the inset of Fig.~\ref{Figure_Asite}(a).

We consider experimentally realized parameters as discussed in the main text. The eigenvalue spectrum of the device, consisting of 28 sites, is shown in Fig.~\ref{Figure_Asite}(a). 
Unlike, Fig.~3(a) in the main text, there is no lattice edge mode in the band gap.
When light is coupled to the boundary site, it remains strongly localized in the linear regime. At weak nonlinearity, localization at the boundary waveguide is not destroyed, as evident by the overlap calculation in Fig.~S3(b). As mentioned previously, flat band modes in our rhombic lattice are localized on the B and C sites of a unit cell with equal intensity and opposite phases. Hence, the flat band modes can not be excited by launching light at the boundary site coupled to the edge A site. 
Additionally, the formation of discrete breathers with oscillating intensity along the propagation distance does not occur in this case.
At larger nonlinearity, relatively stronger detuning of the boundary site causes excitation of the lower dispersive band (Fig.~S3(b)), and hence, light tunnels to the lattice.
}



\begin{thebibliography}{43}%
\makeatletter
\providecommand \@ifxundefined [1]{%
 \@ifx{#1\undefined}
}%
\providecommand \@ifnum [1]{%
 \ifnum #1\expandafter \@firstoftwo
 \else \expandafter \@secondoftwo
 \fi
}%
\providecommand \@ifx [1]{%
 \ifx #1\expandafter \@firstoftwo
 \else \expandafter \@secondoftwo
 \fi
}%
\providecommand \natexlab [1]{#1}%
\providecommand \enquote  [1]{``#1''}%
\providecommand \bibnamefont  [1]{#1}%
\providecommand \bibfnamefont [1]{#1}%
\providecommand \citenamefont [1]{#1}%
\providecommand \href@noop [0]{\@secondoftwo}%
\providecommand \href [0]{\begingroup \@sanitize@url \@href}%
\providecommand \@href[1]{\@@startlink{#1}\@@href}%
\providecommand \@@href[1]{\endgroup#1\@@endlink}%
\providecommand \@sanitize@url [0]{\catcode `\\12\catcode `\$12\catcode `\&12\catcode `\#12\catcode `\^12\catcode `\_12\catcode `\%12\relax}%
\providecommand \@@startlink[1]{}%
\providecommand \@@endlink[0]{}%
\providecommand \url  [0]{\begingroup\@sanitize@url \@url }%
\providecommand \@url [1]{\endgroup\@href {#1}{\urlprefix }}%
\providecommand \urlprefix  [0]{URL }%
\providecommand \Eprint [0]{\href }%
\providecommand \doibase [0]{http://dx.doi.org/}%
\providecommand \selectlanguage [0]{\@gobble}%
\providecommand \bibinfo  [0]{\@secondoftwo}%
\providecommand \bibfield  [0]{\@secondoftwo}%
\providecommand \translation [1]{[#1]}%
\providecommand \BibitemOpen [0]{}%
\providecommand \bibitemStop [0]{}%
\providecommand \bibitemNoStop [0]{.\EOS\space}%
\providecommand \EOS [0]{\spacefactor3000\relax}%
\providecommand \BibitemShut  [1]{\csname bibitem#1\endcsname}%
\let\auto@bib@innerbib\@empty
\bibitem [{\citenamefont {Christodoulides}\ \emph {et~al.}(2003)\citenamefont {Christodoulides}, \citenamefont {Lederer},\ and\ \citenamefont {Silberberg}}]{christodoulides2003discretizing}%
  \BibitemOpen
  \bibfield  {author} {\bibinfo {author} {\bibfnamefont {Demetrios~N}\ \bibnamefont {Christodoulides}}, \bibinfo {author} {\bibfnamefont {Falk}\ \bibnamefont {Lederer}}, \ and\ \bibinfo {author} {\bibfnamefont {Yaron}\ \bibnamefont {Silberberg}},\ }\bibfield  {title} {\enquote {\bibinfo {title} {Discretizing light behaviour in linear and nonlinear waveguide lattices},}\ }\href {\doibase 10.1038/nature01936} {\bibfield  {journal} {\bibinfo  {journal} {Nature}\ }\textbf {\bibinfo {volume} {424}},\ \bibinfo {pages} {817--823} (\bibinfo {year} {2003})}\BibitemShut {NoStop}%
\bibitem [{\citenamefont {Garanovich}\ \emph {et~al.}(2012)\citenamefont {Garanovich}, \citenamefont {Longhi}, \citenamefont {Sukhorukov},\ and\ \citenamefont {Kivshar}}]{Garanovich2012light}%
  \BibitemOpen
  \bibfield  {author} {\bibinfo {author} {\bibfnamefont {Ivan~L.}\ \bibnamefont {Garanovich}}, \bibinfo {author} {\bibfnamefont {Stefano}\ \bibnamefont {Longhi}}, \bibinfo {author} {\bibfnamefont {Andrey~A.}\ \bibnamefont {Sukhorukov}}, \ and\ \bibinfo {author} {\bibfnamefont {Yuri~S.}\ \bibnamefont {Kivshar}},\ }\bibfield  {title} {\enquote {\bibinfo {title} {Light propagation and localization in modulated photonic lattices and waveguides},}\ }\href {\doibase 10.1016/j.physrep.2012.03.005} {\bibfield  {journal} {\bibinfo  {journal} {Phys. Rep.}\ }\textbf {\bibinfo {volume} {518}},\ \bibinfo {pages} {1--79} (\bibinfo {year} {2012})}\BibitemShut {NoStop}%
\bibitem [{\citenamefont {Longhi}(2009)}]{Longhi2009quantum}%
  \BibitemOpen
  \bibfield  {author} {\bibinfo {author} {\bibfnamefont {S.}~\bibnamefont {Longhi}},\ }\bibfield  {title} {\enquote {\bibinfo {title} {Quantum-optical analogies using photonic structures},}\ }\href {\doibase 10.1002/lpor.200810055} {\bibfield  {journal} {\bibinfo  {journal} {Laser \& Photon. Rev.}\ }\textbf {\bibinfo {volume} {3}},\ \bibinfo {pages} {243--261} (\bibinfo {year} {2009})}\BibitemShut {NoStop}%
\bibitem [{\citenamefont {Schwartz}\ \emph {et~al.}(2007)\citenamefont {Schwartz}, \citenamefont {Bartal}, \citenamefont {Fishman},\ and\ \citenamefont {Segev}}]{schwartz2007transport}%
  \BibitemOpen
  \bibfield  {author} {\bibinfo {author} {\bibfnamefont {Tal}\ \bibnamefont {Schwartz}}, \bibinfo {author} {\bibfnamefont {Guy}\ \bibnamefont {Bartal}}, \bibinfo {author} {\bibfnamefont {Shmuel}\ \bibnamefont {Fishman}}, \ and\ \bibinfo {author} {\bibfnamefont {Mordechai}\ \bibnamefont {Segev}},\ }\bibfield  {title} {\enquote {\bibinfo {title} {Transport and {Anderson} localization in disordered two-dimensional photonic lattices},}\ }\href {\doibase 10.1038/nature05623} {\bibfield  {journal} {\bibinfo  {journal} {Nature}\ }\textbf {\bibinfo {volume} {446}},\ \bibinfo {pages} {52--55} (\bibinfo {year} {2007})}\BibitemShut {NoStop}%
\bibitem [{\citenamefont {Segev}\ \emph {et~al.}(2013)\citenamefont {Segev}, \citenamefont {Silberberg},\ and\ \citenamefont {Christodoulides}}]{segev2013anderson}%
  \BibitemOpen
  \bibfield  {author} {\bibinfo {author} {\bibfnamefont {Mordechai}\ \bibnamefont {Segev}}, \bibinfo {author} {\bibfnamefont {Yaron}\ \bibnamefont {Silberberg}}, \ and\ \bibinfo {author} {\bibfnamefont {Demetrios~N}\ \bibnamefont {Christodoulides}},\ }\bibfield  {title} {\enquote {\bibinfo {title} {{Anderson} localization of light},}\ }\href {\doibase 10.1038/nphoton.2013.30} {\bibfield  {journal} {\bibinfo  {journal} {Nat. Photon.}\ }\textbf {\bibinfo {volume} {7}},\ \bibinfo {pages} {197--204} (\bibinfo {year} {2013})}\BibitemShut {NoStop}%
\bibitem [{\citenamefont {Christodoulides}\ and\ \citenamefont {Joseph}(1988)}]{Christodoulides1988}%
  \BibitemOpen
  \bibfield  {author} {\bibinfo {author} {\bibfnamefont {D.~N.}\ \bibnamefont {Christodoulides}}\ and\ \bibinfo {author} {\bibfnamefont {R.~I.}\ \bibnamefont {Joseph}},\ }\bibfield  {title} {\enquote {\bibinfo {title} {Discrete self-focusing in nonlinear arrays of coupled waveguides},}\ }\href {\doibase 10.1364/OL.13.000794} {\bibfield  {journal} {\bibinfo  {journal} {Opt. Lett.}\ }\textbf {\bibinfo {volume} {13}},\ \bibinfo {pages} {794--796} (\bibinfo {year} {1988})}\BibitemShut {NoStop}%
\bibitem [{\citenamefont {Segev}\ \emph {et~al.}(1992)\citenamefont {Segev}, \citenamefont {Crosignani}, \citenamefont {Yariv},\ and\ \citenamefont {Fischer}}]{segev1992spatial}%
  \BibitemOpen
  \bibfield  {author} {\bibinfo {author} {\bibfnamefont {Mordechai}\ \bibnamefont {Segev}}, \bibinfo {author} {\bibfnamefont {Bruno}\ \bibnamefont {Crosignani}}, \bibinfo {author} {\bibfnamefont {Amnon}\ \bibnamefont {Yariv}}, \ and\ \bibinfo {author} {\bibfnamefont {Baruch}\ \bibnamefont {Fischer}},\ }\bibfield  {title} {\enquote {\bibinfo {title} {Spatial solitons in photorefractive media},}\ }\href {\doibase 10.1103/PhysRevLett.68.923} {\bibfield  {journal} {\bibinfo  {journal} {Phys. Rev. Lett.}\ }\textbf {\bibinfo {volume} {68}},\ \bibinfo {pages} {923} (\bibinfo {year} {1992})}\BibitemShut {NoStop}%
\bibitem [{\citenamefont {Eisenberg}\ \emph {et~al.}(1998)\citenamefont {Eisenberg}, \citenamefont {Silberberg}, \citenamefont {Morandotti}, \citenamefont {Boyd},\ and\ \citenamefont {Aitchison}}]{eisenberg1998discrete}%
  \BibitemOpen
  \bibfield  {author} {\bibinfo {author} {\bibfnamefont {H.~S.}\ \bibnamefont {Eisenberg}}, \bibinfo {author} {\bibfnamefont {Yaron}\ \bibnamefont {Silberberg}}, \bibinfo {author} {\bibfnamefont {R}~\bibnamefont {Morandotti}}, \bibinfo {author} {\bibfnamefont {A.~R.}\ \bibnamefont {Boyd}}, \ and\ \bibinfo {author} {\bibfnamefont {J.~S.}\ \bibnamefont {Aitchison}},\ }\bibfield  {title} {\enquote {\bibinfo {title} {Discrete spatial optical solitons in waveguide arrays},}\ }\href {\doibase 10.1103/PhysRevLett.81.3383} {\bibfield  {journal} {\bibinfo  {journal} {Phys. Rev. Lett.}\ }\textbf {\bibinfo {volume} {81}},\ \bibinfo {pages} {3383} (\bibinfo {year} {1998})}\BibitemShut {NoStop}%
\bibitem [{\citenamefont {Lederer}\ \emph {et~al.}(2008)\citenamefont {Lederer}, \citenamefont {Stegeman}, \citenamefont {Christodoulides}, \citenamefont {Assanto}, \citenamefont {Segev},\ and\ \citenamefont {Silberberg}}]{lederer2008discrete}%
  \BibitemOpen
  \bibfield  {author} {\bibinfo {author} {\bibfnamefont {Falk}\ \bibnamefont {Lederer}}, \bibinfo {author} {\bibfnamefont {George~I}\ \bibnamefont {Stegeman}}, \bibinfo {author} {\bibfnamefont {Demetri~N}\ \bibnamefont {Christodoulides}}, \bibinfo {author} {\bibfnamefont {Gaetano}\ \bibnamefont {Assanto}}, \bibinfo {author} {\bibfnamefont {Moti}\ \bibnamefont {Segev}}, \ and\ \bibinfo {author} {\bibfnamefont {Yaron}\ \bibnamefont {Silberberg}},\ }\bibfield  {title} {\enquote {\bibinfo {title} {Discrete solitons in optics},}\ }\href {\doibase 10.1016/j.physrep.2008.04.004} {\bibfield  {journal} {\bibinfo  {journal} {Phys. Rep.}\ }\textbf {\bibinfo {volume} {463}},\ \bibinfo {pages} {1--126} (\bibinfo {year} {2008})}\BibitemShut {NoStop}%
\bibitem [{\citenamefont {Flach}\ and\ \citenamefont {Willis}(1998)}]{flach1998discrete}%
  \BibitemOpen
  \bibfield  {author} {\bibinfo {author} {\bibfnamefont {Sergej}\ \bibnamefont {Flach}}\ and\ \bibinfo {author} {\bibfnamefont {Charles~R}\ \bibnamefont {Willis}},\ }\bibfield  {title} {\enquote {\bibinfo {title} {Discrete breathers},}\ }\href {\doibase 10.1016/S0370-1573(97)00068-9} {\bibfield  {journal} {\bibinfo  {journal} {Phys. Rep.}\ }\textbf {\bibinfo {volume} {295}},\ \bibinfo {pages} {181--264} (\bibinfo {year} {1998})}\BibitemShut {NoStop}%
\bibitem [{\citenamefont {Kopidakis}\ and\ \citenamefont {Aubry}(2000)}]{kopidakis2000discrete}%
  \BibitemOpen
  \bibfield  {author} {\bibinfo {author} {\bibfnamefont {G}~\bibnamefont {Kopidakis}}\ and\ \bibinfo {author} {\bibfnamefont {S}~\bibnamefont {Aubry}},\ }\bibfield  {title} {\enquote {\bibinfo {title} {Discrete breathers and delocalization in nonlinear disordered systems},}\ }\href {\doibase 10.1103/PhysRevLett.84.3236} {\bibfield  {journal} {\bibinfo  {journal} {Phys. Rev. Lett.}\ }\textbf {\bibinfo {volume} {84}},\ \bibinfo {pages} {3236} (\bibinfo {year} {2000})}\BibitemShut {NoStop}%
\bibitem [{\citenamefont {Mandelik}\ \emph {et~al.}(2003)\citenamefont {Mandelik}, \citenamefont {Eisenberg}, \citenamefont {Silberberg}, \citenamefont {Morandotti},\ and\ \citenamefont {Aitchison}}]{mandelik2003observation}%
  \BibitemOpen
  \bibfield  {author} {\bibinfo {author} {\bibfnamefont {Diana}\ \bibnamefont {Mandelik}}, \bibinfo {author} {\bibfnamefont {HS}~\bibnamefont {Eisenberg}}, \bibinfo {author} {\bibfnamefont {Yaron}\ \bibnamefont {Silberberg}}, \bibinfo {author} {\bibfnamefont {R}~\bibnamefont {Morandotti}}, \ and\ \bibinfo {author} {\bibfnamefont {JS}~\bibnamefont {Aitchison}},\ }\bibfield  {title} {\enquote {\bibinfo {title} {Observation of mutually trapped multiband optical breathers in waveguide arrays},}\ }\href {\doibase 10.1103/PhysRevLett.90.253902} {\bibfield  {journal} {\bibinfo  {journal} {Phys. Rev. Lett.}\ }\textbf {\bibinfo {volume} {90}},\ \bibinfo {pages} {253902} (\bibinfo {year} {2003})}\BibitemShut {NoStop}%
\bibitem [{\citenamefont {Shit}\ \emph {et~al.}(2025)\citenamefont {Shit}, \citenamefont {Hui}, \citenamefont {Di~Liberto}, \citenamefont {Sen},\ and\ \citenamefont {Mukherjee}}]{shit2024probing}%
  \BibitemOpen
  \bibfield  {author} {\bibinfo {author} {\bibfnamefont {Trideb}\ \bibnamefont {Shit}}, \bibinfo {author} {\bibfnamefont {Rishav}\ \bibnamefont {Hui}}, \bibinfo {author} {\bibfnamefont {Marco}\ \bibnamefont {Di~Liberto}}, \bibinfo {author} {\bibfnamefont {Diptiman}\ \bibnamefont {Sen}}, \ and\ \bibinfo {author} {\bibfnamefont {Sebabrata}\ \bibnamefont {Mukherjee}},\ }\bibfield  {title} {\enquote {\bibinfo {title} {Intensity correlation measurement to simulate two-body bound states in the continuum and probe nonlinear discrete breathers},}\ }\href {\doibase 10.1103/PhysRevA.111.053515} {\bibfield  {journal} {\bibinfo  {journal} {Phys. Rev. A}\ }\textbf {\bibinfo {volume} {111}},\ \bibinfo {pages} {053515} (\bibinfo {year} {2025})}\BibitemShut {NoStop}%
\bibitem [{\citenamefont {Geniet}\ and\ \citenamefont {Leon}(2002)}]{F_Geniet_PRL}%
  \BibitemOpen
  \bibfield  {author} {\bibinfo {author} {\bibfnamefont {F.}~\bibnamefont {Geniet}}\ and\ \bibinfo {author} {\bibfnamefont {J.}~\bibnamefont {Leon}},\ }\bibfield  {title} {\enquote {\bibinfo {title} {Energy transmission in the forbidden band gap of a nonlinear chain},}\ }\href {\doibase 10.1103/PhysRevLett.89.134102} {\bibfield  {journal} {\bibinfo  {journal} {Phys. Rev. Lett.}\ }\textbf {\bibinfo {volume} {89}},\ \bibinfo {pages} {134102} (\bibinfo {year} {2002})}\BibitemShut {NoStop}%
\bibitem [{\citenamefont {Khomeriki}(2004)}]{Ramaz}%
  \BibitemOpen
  \bibfield  {author} {\bibinfo {author} {\bibfnamefont {Ramaz}\ \bibnamefont {Khomeriki}},\ }\bibfield  {title} {\enquote {\bibinfo {title} {Nonlinear band gap transmission in optical waveguide arrays},}\ }\href {\doibase 10.1103/PhysRevLett.92.063905} {\bibfield  {journal} {\bibinfo  {journal} {Phys. Rev. Lett.}\ }\textbf {\bibinfo {volume} {92}},\ \bibinfo {pages} {063905} (\bibinfo {year} {2004})}\BibitemShut {NoStop}%
\bibitem [{\citenamefont {Susanto}\ and\ \citenamefont {Karjanto}(2008)}]{susanto2008calculated}%
  \BibitemOpen
  \bibfield  {author} {\bibinfo {author} {\bibfnamefont {H}~\bibnamefont {Susanto}}\ and\ \bibinfo {author} {\bibfnamefont {N}~\bibnamefont {Karjanto}},\ }\bibfield  {title} {\enquote {\bibinfo {title} {Calculated threshold of supratransmission phenomena in waveguide arrays with saturable nonlinearity},}\ }\href {\doibase 10.1142/S0218863508004147} {\bibfield  {journal} {\bibinfo  {journal} {J. Nonlinear Opt. Phys. \& Mat.}\ }\textbf {\bibinfo {volume} {17}},\ \bibinfo {pages} {159--165} (\bibinfo {year} {2008})}\BibitemShut {NoStop}%
\bibitem [{\citenamefont {Motcheyo}\ \emph {et~al.}(2017)\citenamefont {Motcheyo}, \citenamefont {Tchameu}, \citenamefont {Siewe},\ and\ \citenamefont {Tchawoua}}]{motcheyo2017homoclinic}%
  \BibitemOpen
  \bibfield  {author} {\bibinfo {author} {\bibfnamefont {AB~Togueu}\ \bibnamefont {Motcheyo}}, \bibinfo {author} {\bibfnamefont {JD~Tchinang}\ \bibnamefont {Tchameu}}, \bibinfo {author} {\bibfnamefont {M~Siewe}\ \bibnamefont {Siewe}}, \ and\ \bibinfo {author} {\bibfnamefont {Cl{\'e}ment}\ \bibnamefont {Tchawoua}},\ }\bibfield  {title} {\enquote {\bibinfo {title} {Homoclinic nonlinear band gap transmission threshold in discrete optical waveguide arrays},}\ }\href {\doibase 10.1016/j.cnsns.2017.02.001} {\bibfield  {journal} {\bibinfo  {journal} {Communications in Nonlinear Science and Numerical Simulation}\ }\textbf {\bibinfo {volume} {50}},\ \bibinfo {pages} {29--34} (\bibinfo {year} {2017})}\BibitemShut {NoStop}%
\bibitem [{\citenamefont {Zakharov}(2023)}]{zakharov2023effect}%
  \BibitemOpen
  \bibfield  {author} {\bibinfo {author} {\bibfnamefont {P.}~\bibnamefont {Zakharov}},\ }\bibfield  {title} {\enquote {\bibinfo {title} {The effect of nonlinear supratransmission in discrete structures: A review},}\ }\href {\doibase 10.20537/2076-7633-2023-15-3-599-617} {\bibfield  {journal} {\bibinfo  {journal} {Comput. Res. Model}\ }\textbf {\bibinfo {volume} {15}},\ \bibinfo {pages} {599--617} (\bibinfo {year} {2023})}\BibitemShut {NoStop}%
\bibitem [{\citenamefont {Susanto}\ \emph {et~al.}(2023)\citenamefont {Susanto}, \citenamefont {Lazarides},\ and\ \citenamefont {Kourakis}}]{susanto2023surge}%
  \BibitemOpen
  \bibfield  {author} {\bibinfo {author} {\bibfnamefont {H}~\bibnamefont {Susanto}}, \bibinfo {author} {\bibfnamefont {N}~\bibnamefont {Lazarides}}, \ and\ \bibinfo {author} {\bibfnamefont {I}~\bibnamefont {Kourakis}},\ }\bibfield  {title} {\enquote {\bibinfo {title} {Surge of power transmission in flat and nearly flat band lattices},}\ }\href {\doibase 10.1103/PhysRevE.108.L052201} {\bibfield  {journal} {\bibinfo  {journal} {Phys. Rev. E}\ }\textbf {\bibinfo {volume} {108}},\ \bibinfo {pages} {L052201} (\bibinfo {year} {2023})}\BibitemShut {NoStop}%
\bibitem [{\citenamefont {Macías-Díaz}\ and\ \citenamefont {Puri}(2007)}]{MACIASDIAZ2007447}%
  \BibitemOpen
  \bibfield  {author} {\bibinfo {author} {\bibfnamefont {J.E.}\ \bibnamefont {Macías-Díaz}}\ and\ \bibinfo {author} {\bibfnamefont {A.}~\bibnamefont {Puri}},\ }\bibfield  {title} {\enquote {\bibinfo {title} {An application of nonlinear supratransmission to the propagation of binary signals in weakly damped, mechanical systems of coupled oscillators},}\ }\href {\doibase 10.1016/j.physleta.2007.03.076} {\bibfield  {journal} {\bibinfo  {journal} {Phys. Lett. A}\ }\textbf {\bibinfo {volume} {366}},\ \bibinfo {pages} {447--450} (\bibinfo {year} {2007})}\BibitemShut {NoStop}%
\bibitem [{\citenamefont {Togueu~Motcheyo}\ and\ \citenamefont {Macías-Díaz}(2023)}]{Motcheyo}%
  \BibitemOpen
  \bibfield  {author} {\bibinfo {author} {\bibfnamefont {Alain~Bertrand}\ \bibnamefont {Togueu~Motcheyo}}\ and\ \bibinfo {author} {\bibfnamefont {J.}~\bibnamefont {Macías-Díaz}},\ }\bibfield  {title} {\enquote {\bibinfo {title} {Nonlinear bandgap transmission with zero frequency in a cross-stitch lattice},}\ }\href {\doibase 10.1016/j.chaos.2023.113349} {\bibfield  {journal} {\bibinfo  {journal} {Chaos Solitons and Fractals}\ }\textbf {\bibinfo {volume} {170}},\ \bibinfo {pages} {113349} (\bibinfo {year} {2023})}\BibitemShut {NoStop}%
\bibitem [{\citenamefont {Tasaki}(2008)}]{tasaki2008hubbard}%
  \BibitemOpen
  \bibfield  {author} {\bibinfo {author} {\bibfnamefont {H}~\bibnamefont {Tasaki}},\ }\bibfield  {title} {\enquote {\bibinfo {title} {Hubbard model and the origin of ferromagnetism},}\ }\href {\doibase 10.1140/epjb/e2008-00113-2} {\bibfield  {journal} {\bibinfo  {journal} {Eur. Phys. J. B}\ }\textbf {\bibinfo {volume} {64}},\ \bibinfo {pages} {365--372} (\bibinfo {year} {2008})}\BibitemShut {NoStop}%
\bibitem [{\citenamefont {Leykam}\ and\ \citenamefont {Flach}(2018)}]{leykam2018perspective}%
  \BibitemOpen
  \bibfield  {author} {\bibinfo {author} {\bibfnamefont {Daniel}\ \bibnamefont {Leykam}}\ and\ \bibinfo {author} {\bibfnamefont {Sergej}\ \bibnamefont {Flach}},\ }\bibfield  {title} {\enquote {\bibinfo {title} {Perspective: photonic flatbands},}\ }\href {\doibase 10.1063/1.5034365} {\bibfield  {journal} {\bibinfo  {journal} {APL Photon.}\ }\textbf {\bibinfo {volume} {3}},\ \bibinfo {pages} {070901} (\bibinfo {year} {2018})}\BibitemShut {NoStop}%
\bibitem [{\citenamefont {Mukherjee}\ \emph {et~al.}(2015)\citenamefont {Mukherjee}, \citenamefont {Spracklen}, \citenamefont {Choudhury}, \citenamefont {Goldman}, \citenamefont {\"Ohberg}, \citenamefont {Andersson},\ and\ \citenamefont {Thomson}}]{FB_PRL_SM_Lieb}%
  \BibitemOpen
  \bibfield  {author} {\bibinfo {author} {\bibfnamefont {Sebabrata}\ \bibnamefont {Mukherjee}}, \bibinfo {author} {\bibfnamefont {Alexander}\ \bibnamefont {Spracklen}}, \bibinfo {author} {\bibfnamefont {Debaditya}\ \bibnamefont {Choudhury}}, \bibinfo {author} {\bibfnamefont {Nathan}\ \bibnamefont {Goldman}}, \bibinfo {author} {\bibfnamefont {Patrik}\ \bibnamefont {\"Ohberg}}, \bibinfo {author} {\bibfnamefont {Erika}\ \bibnamefont {Andersson}}, \ and\ \bibinfo {author} {\bibfnamefont {Robert~R.}\ \bibnamefont {Thomson}},\ }\bibfield  {title} {\enquote {\bibinfo {title} {Observation of a localized flat-band state in a photonic {Lieb} lattice},}\ }\href {\doibase 10.1103/PhysRevLett.114.245504} {\bibfield  {journal} {\bibinfo  {journal} {Phys. Rev. Lett.}\ }\textbf {\bibinfo {volume} {114}},\ \bibinfo {pages} {245504} (\bibinfo {year} {2015})}\BibitemShut {NoStop}%
\bibitem [{\citenamefont {Vicencio}\ \emph {et~al.}(2015)\citenamefont {Vicencio}, \citenamefont {Cantillano}, \citenamefont {Morales-Inostroza}, \citenamefont {Real}, \citenamefont {Mej{\'\i}a-Cort{\'e}s}, \citenamefont {Weimann}, \citenamefont {Szameit},\ and\ \citenamefont {Molina}}]{vicencio2015observation}%
  \BibitemOpen
  \bibfield  {author} {\bibinfo {author} {\bibfnamefont {Rodrigo~A}\ \bibnamefont {Vicencio}}, \bibinfo {author} {\bibfnamefont {Camilo}\ \bibnamefont {Cantillano}}, \bibinfo {author} {\bibfnamefont {Luis}\ \bibnamefont {Morales-Inostroza}}, \bibinfo {author} {\bibfnamefont {Basti{\'a}n}\ \bibnamefont {Real}}, \bibinfo {author} {\bibfnamefont {Cristian}\ \bibnamefont {Mej{\'\i}a-Cort{\'e}s}}, \bibinfo {author} {\bibfnamefont {Steffen}\ \bibnamefont {Weimann}}, \bibinfo {author} {\bibfnamefont {Alexander}\ \bibnamefont {Szameit}}, \ and\ \bibinfo {author} {\bibfnamefont {Mario~I}\ \bibnamefont {Molina}},\ }\bibfield  {title} {\enquote {\bibinfo {title} {Observation of localized states in {Lieb} photonic lattices},}\ }\href {\doibase 10.1103/PhysRevLett.114.245503} {\bibfield  {journal} {\bibinfo  {journal} {Phys. Rev. Lett.}\ }\textbf {\bibinfo {volume} {114}},\ \bibinfo {pages} {245503} (\bibinfo {year} {2015})}\BibitemShut {NoStop}%
\bibitem [{\citenamefont {Mukherjee}\ and\ \citenamefont {Thomson}(2015)}]{RhombicFB_SM}%
  \BibitemOpen
  \bibfield  {author} {\bibinfo {author} {\bibfnamefont {Sebabrata}\ \bibnamefont {Mukherjee}}\ and\ \bibinfo {author} {\bibfnamefont {Robert~R.}\ \bibnamefont {Thomson}},\ }\bibfield  {title} {\enquote {\bibinfo {title} {Observation of localized flat-band modes in a quasi-one-dimensional photonic rhombic lattice},}\ }\href {\doibase 10.1364/OL.40.005443} {\bibfield  {journal} {\bibinfo  {journal} {Opt. Lett.}\ }\textbf {\bibinfo {volume} {40}},\ \bibinfo {pages} {5443--5446} (\bibinfo {year} {2015})}\BibitemShut {NoStop}%
\bibitem [{\citenamefont {Guzmán-Silva}\ \emph {et~al.}(2014)\citenamefont {Guzmán-Silva}, \citenamefont {Mejía-Cortés}, \citenamefont {Bandres}, \citenamefont {Rechtsman}, \citenamefont {Weimann}, \citenamefont {Nolte}, \citenamefont {Segev}, \citenamefont {Szameit},\ and\ \citenamefont {Vicencio}}]{FB_Lieb_IOP}%
  \BibitemOpen
  \bibfield  {author} {\bibinfo {author} {\bibfnamefont {D}~\bibnamefont {Guzmán-Silva}}, \bibinfo {author} {\bibfnamefont {C}~\bibnamefont {Mejía-Cortés}}, \bibinfo {author} {\bibfnamefont {M~A}\ \bibnamefont {Bandres}}, \bibinfo {author} {\bibfnamefont {M~C}\ \bibnamefont {Rechtsman}}, \bibinfo {author} {\bibfnamefont {S}~\bibnamefont {Weimann}}, \bibinfo {author} {\bibfnamefont {S}~\bibnamefont {Nolte}}, \bibinfo {author} {\bibfnamefont {M}~\bibnamefont {Segev}}, \bibinfo {author} {\bibfnamefont {A}~\bibnamefont {Szameit}}, \ and\ \bibinfo {author} {\bibfnamefont {R~A}\ \bibnamefont {Vicencio}},\ }\bibfield  {title} {\enquote {\bibinfo {title} {Experimental observation of bulk and edge transport in photonic {Lieb} lattices},}\ }\href {\doibase 10.1088/1367-2630/16/6/063061} {\bibfield  {journal} {\bibinfo  {journal} {New J. Phys.}\ }\textbf {\bibinfo {volume} {16}},\ \bibinfo {pages} {063061} (\bibinfo {year} {2014})}\BibitemShut {NoStop}%
\bibitem [{\citenamefont {Mukherjee}\ \emph {et~al.}(2018)\citenamefont {Mukherjee}, \citenamefont {Di~Liberto}, \citenamefont {{\"O}hberg}, \citenamefont {Thomson},\ and\ \citenamefont {Goldman}}]{mukherjee2018experimental}%
  \BibitemOpen
  \bibfield  {author} {\bibinfo {author} {\bibfnamefont {Sebabrata}\ \bibnamefont {Mukherjee}}, \bibinfo {author} {\bibfnamefont {Marco}\ \bibnamefont {Di~Liberto}}, \bibinfo {author} {\bibfnamefont {Patrik}\ \bibnamefont {{\"O}hberg}}, \bibinfo {author} {\bibfnamefont {Robert~R}\ \bibnamefont {Thomson}}, \ and\ \bibinfo {author} {\bibfnamefont {Nathan}\ \bibnamefont {Goldman}},\ }\bibfield  {title} {\enquote {\bibinfo {title} {Experimental observation of {Aharonov-Bohm} cages in photonic lattices},}\ }\href {\doibase doi.org/10.1103/PhysRevLett.121.075502} {\bibfield  {journal} {\bibinfo  {journal} {Phys. Rev. Lett.}\ }\textbf {\bibinfo {volume} {121}},\ \bibinfo {pages} {075502} (\bibinfo {year} {2018})}\BibitemShut {NoStop}%
\bibitem [{\citenamefont {Xia}\ \emph {et~al.}(2016)\citenamefont {Xia}, \citenamefont {Hu}, \citenamefont {Song}, \citenamefont {Zong}, \citenamefont {Tang},\ and\ \citenamefont {Chen}}]{xia2016demonstration}%
  \BibitemOpen
  \bibfield  {author} {\bibinfo {author} {\bibfnamefont {Shiqiang}\ \bibnamefont {Xia}}, \bibinfo {author} {\bibfnamefont {Yi}~\bibnamefont {Hu}}, \bibinfo {author} {\bibfnamefont {Daohong}\ \bibnamefont {Song}}, \bibinfo {author} {\bibfnamefont {Yuanyuan}\ \bibnamefont {Zong}}, \bibinfo {author} {\bibfnamefont {Liqin}\ \bibnamefont {Tang}}, \ and\ \bibinfo {author} {\bibfnamefont {Zhigang}\ \bibnamefont {Chen}},\ }\bibfield  {title} {\enquote {\bibinfo {title} {Demonstration of flat-band image transmission in optically induced {Lieb} photonic lattices},}\ }\href {\doibase 10.1364/OL.41.001435} {\bibfield  {journal} {\bibinfo  {journal} {Opt. Lett.}\ }\textbf {\bibinfo {volume} {41}},\ \bibinfo {pages} {1435--1438} (\bibinfo {year} {2016})}\BibitemShut {NoStop}%
\bibitem [{\citenamefont {Taie}\ \emph {et~al.}(2015)\citenamefont {Taie}, \citenamefont {Ozawa}, \citenamefont {Ichinose}, \citenamefont {Nishio}, \citenamefont {Nakajima},\ and\ \citenamefont {Takahashi}}]{taie2015coherent}%
  \BibitemOpen
  \bibfield  {author} {\bibinfo {author} {\bibfnamefont {Shintaro}\ \bibnamefont {Taie}}, \bibinfo {author} {\bibfnamefont {Hideki}\ \bibnamefont {Ozawa}}, \bibinfo {author} {\bibfnamefont {Tomohiro}\ \bibnamefont {Ichinose}}, \bibinfo {author} {\bibfnamefont {Takuei}\ \bibnamefont {Nishio}}, \bibinfo {author} {\bibfnamefont {Shuta}\ \bibnamefont {Nakajima}}, \ and\ \bibinfo {author} {\bibfnamefont {Yoshiro}\ \bibnamefont {Takahashi}},\ }\bibfield  {title} {\enquote {\bibinfo {title} {Coherent driving and freezing of bosonic matter wave in an optical {Lieb} lattice},}\ }\href {\doibase 10.1126/sciadv.1500854} {\bibfield  {journal} {\bibinfo  {journal} {Science Advances}\ }\textbf {\bibinfo {volume} {1}},\ \bibinfo {pages} {e1500854} (\bibinfo {year} {2015})}\BibitemShut {NoStop}%
\bibitem [{\citenamefont {Zeng}\ \emph {et~al.}(2024)\citenamefont {Zeng}, \citenamefont {Shi}, \citenamefont {Mao}, \citenamefont {Wu}, \citenamefont {Xie}, \citenamefont {Yuan}, \citenamefont {Zhang}, \citenamefont {Dai}, \citenamefont {Chen},\ and\ \citenamefont {Pan}}]{zeng2024transition}%
  \BibitemOpen
  \bibfield  {author} {\bibinfo {author} {\bibfnamefont {Chao}\ \bibnamefont {Zeng}}, \bibinfo {author} {\bibfnamefont {Yue-Ran}\ \bibnamefont {Shi}}, \bibinfo {author} {\bibfnamefont {Yi-Yi}\ \bibnamefont {Mao}}, \bibinfo {author} {\bibfnamefont {Fei-Fei}\ \bibnamefont {Wu}}, \bibinfo {author} {\bibfnamefont {Yan-Jun}\ \bibnamefont {Xie}}, \bibinfo {author} {\bibfnamefont {Tao}\ \bibnamefont {Yuan}}, \bibinfo {author} {\bibfnamefont {Wei}\ \bibnamefont {Zhang}}, \bibinfo {author} {\bibfnamefont {Han-Ning}\ \bibnamefont {Dai}}, \bibinfo {author} {\bibfnamefont {Yu-Ao}\ \bibnamefont {Chen}}, \ and\ \bibinfo {author} {\bibfnamefont {Jian-Wei}\ \bibnamefont {Pan}},\ }\bibfield  {title} {\enquote {\bibinfo {title} {Transition from flat-band localization to {Anderson} localization in a one-dimensional tasaki lattice},}\ }\href {\doibase 10.1103/PhysRevLett.132.063401} {\bibfield  {journal} {\bibinfo  {journal} {Phys. Rev. Lett.}\ }\textbf {\bibinfo {volume} {132}},\ \bibinfo {pages} {063401} (\bibinfo {year}
  {2024})}\BibitemShut {NoStop}%
\bibitem [{\citenamefont {Schulz}\ \emph {et~al.}(2017)\citenamefont {Schulz}, \citenamefont {Upham}, \citenamefont {O’Faolain},\ and\ \citenamefont {Boyd}}]{schulz2017photonic}%
  \BibitemOpen
  \bibfield  {author} {\bibinfo {author} {\bibfnamefont {Sebastian~A}\ \bibnamefont {Schulz}}, \bibinfo {author} {\bibfnamefont {Jeremy}\ \bibnamefont {Upham}}, \bibinfo {author} {\bibfnamefont {Liam}\ \bibnamefont {O’Faolain}}, \ and\ \bibinfo {author} {\bibfnamefont {Robert~W}\ \bibnamefont {Boyd}},\ }\bibfield  {title} {\enquote {\bibinfo {title} {Photonic crystal slow light waveguides in a kagome lattice},}\ }\href {\doibase 10.1364/OL.42.003243} {\bibfield  {journal} {\bibinfo  {journal} {Opt. Lett.}\ }\textbf {\bibinfo {volume} {42}},\ \bibinfo {pages} {3243--3246} (\bibinfo {year} {2017})}\BibitemShut {NoStop}%
\bibitem [{\citenamefont {Agrawal}(2000)}]{agrawal2000nonlinear}%
  \BibitemOpen
  \bibfield  {author} {\bibinfo {author} {\bibfnamefont {Govind~P}\ \bibnamefont {Agrawal}},\ }\bibfield  {title} {\enquote {\bibinfo {title} {Nonlinear fiber optics},}\ }in\ \href@noop {} {\emph {\bibinfo {booktitle} {Nonlinear Science at the Dawn of the 21st Century}}}\ (\bibinfo  {publisher} {Springer},\ \bibinfo {year} {2000})\BibitemShut {NoStop}%
\bibitem [{\citenamefont {Ams}\ \emph {et~al.}(2005)\citenamefont {Ams}, \citenamefont {Marshall}, \citenamefont {Spence},\ and\ \citenamefont {Withford}}]{Ams}%
  \BibitemOpen
  \bibfield  {author} {\bibinfo {author} {\bibfnamefont {M.}~\bibnamefont {Ams}}, \bibinfo {author} {\bibfnamefont {G.~D.}\ \bibnamefont {Marshall}}, \bibinfo {author} {\bibfnamefont {D.~J.}\ \bibnamefont {Spence}}, \ and\ \bibinfo {author} {\bibfnamefont {M.~J.}\ \bibnamefont {Withford}},\ }\bibfield  {title} {\enquote {\bibinfo {title} {Slit beam shaping method for femtosecond laser direct-write fabrication of symmetric waveguides in bulk glasses},}\ }\href {\doibase 10.1364/OPEX.13.005676} {\bibfield  {journal} {\bibinfo  {journal} {Opt. Express}\ }\textbf {\bibinfo {volume} {13}},\ \bibinfo {pages} {5676--5681} (\bibinfo {year} {2005})}\BibitemShut {NoStop}%
\bibitem [{\citenamefont {Davis}\ \emph {et~al.}(1996)\citenamefont {Davis}, \citenamefont {Miura}, \citenamefont {Sugimoto},\ and\ \citenamefont {Hirao}}]{Davis}%
  \BibitemOpen
  \bibfield  {author} {\bibinfo {author} {\bibfnamefont {K.~M.}\ \bibnamefont {Davis}}, \bibinfo {author} {\bibfnamefont {K.}~\bibnamefont {Miura}}, \bibinfo {author} {\bibfnamefont {N.}~\bibnamefont {Sugimoto}}, \ and\ \bibinfo {author} {\bibfnamefont {K.}~\bibnamefont {Hirao}},\ }\bibfield  {title} {\enquote {\bibinfo {title} {Writing waveguides in glass with a femtosecond laser},}\ }\href {\doibase 10.1364/OL.21.001729} {\bibfield  {journal} {\bibinfo  {journal} {Opt. Lett.}\ }\textbf {\bibinfo {volume} {21}},\ \bibinfo {pages} {1729--1731} (\bibinfo {year} {1996})}\BibitemShut {NoStop}%
\bibitem [{\citenamefont {Friberg}\ \emph {et~al.}(1987)\citenamefont {Friberg}, \citenamefont {Silberberg}, \citenamefont {Oliver}, \citenamefont {Andrejco}, \citenamefont {Saifi},\ and\ \citenamefont {Smith}}]{friberg1987ultrafast}%
  \BibitemOpen
  \bibfield  {author} {\bibinfo {author} {\bibfnamefont {SR}~\bibnamefont {Friberg}}, \bibinfo {author} {\bibfnamefont {Y}~\bibnamefont {Silberberg}}, \bibinfo {author} {\bibfnamefont {MKr}\ \bibnamefont {Oliver}}, \bibinfo {author} {\bibfnamefont {MJ}~\bibnamefont {Andrejco}}, \bibinfo {author} {\bibfnamefont {MA}~\bibnamefont {Saifi}}, \ and\ \bibinfo {author} {\bibfnamefont {PW}~\bibnamefont {Smith}},\ }\bibfield  {title} {\enquote {\bibinfo {title} {Ultrafast all-optical switching in a dual-core fiber nonlinear coupler},}\ }\href {\doibase 10.1063/1.98762} {\bibfield  {journal} {\bibinfo  {journal} {Appl. Phys. Lett.}\ }\textbf {\bibinfo {volume} {51}},\ \bibinfo {pages} {1135--1137} (\bibinfo {year} {1987})}\BibitemShut {NoStop}%
\bibitem [{\citenamefont {Demetriou}\ \emph {et~al.}(2017)\citenamefont {Demetriou}, \citenamefont {Hewak}, \citenamefont {Ravagli}, \citenamefont {Craig},\ and\ \citenamefont {Kar}}]{demetriou2017nonlinear}%
  \BibitemOpen
  \bibfield  {author} {\bibinfo {author} {\bibfnamefont {Giorgos}\ \bibnamefont {Demetriou}}, \bibinfo {author} {\bibfnamefont {Daniel~W}\ \bibnamefont {Hewak}}, \bibinfo {author} {\bibfnamefont {Andrea}\ \bibnamefont {Ravagli}}, \bibinfo {author} {\bibfnamefont {Chris}\ \bibnamefont {Craig}}, \ and\ \bibinfo {author} {\bibfnamefont {Ajoy}\ \bibnamefont {Kar}},\ }\bibfield  {title} {\enquote {\bibinfo {title} {Nonlinear refractive index of ultrafast laser inscribed waveguides in gallium lanthanum sulphide},}\ }\href@noop {} {\bibfield  {journal} {\bibinfo  {journal} {Appl. Optics}\ }\textbf {\bibinfo {volume} {56}},\ \bibinfo {pages} {5407--5411} (\bibinfo {year} {2017})}\BibitemShut {NoStop}%
\bibitem [{\citenamefont {Rajeevan}\ and\ \citenamefont {Mukherjee}(2025)}]{rajeevan2025nonlinear}%
  \BibitemOpen
  \bibfield  {author} {\bibinfo {author} {\bibfnamefont {Gayathry}\ \bibnamefont {Rajeevan}}\ and\ \bibinfo {author} {\bibfnamefont {Sebabrata}\ \bibnamefont {Mukherjee}},\ }\bibfield  {title} {\enquote {\bibinfo {title} {Nonlinear switch and spatial lattice solitons of photonic s--p orbitals},}\ }\href {\doibase 10.1364/OL.546876} {\bibfield  {journal} {\bibinfo  {journal} {Opt. Lett.}\ }\textbf {\bibinfo {volume} {50}},\ \bibinfo {pages} {297--300} (\bibinfo {year} {2025})}\BibitemShut {NoStop}%
\bibitem [{\citenamefont {Cao}\ \emph {et~al.}(2018)\citenamefont {Cao}, \citenamefont {Fatemi}, \citenamefont {Fang}, \citenamefont {Watanabe}, \citenamefont {Taniguchi}, \citenamefont {Kaxiras},\ and\ \citenamefont {Jarillo-Herrero}}]{cao2018unconventional}%
  \BibitemOpen
  \bibfield  {author} {\bibinfo {author} {\bibfnamefont {Yuan}\ \bibnamefont {Cao}}, \bibinfo {author} {\bibfnamefont {Valla}\ \bibnamefont {Fatemi}}, \bibinfo {author} {\bibfnamefont {Shiang}\ \bibnamefont {Fang}}, \bibinfo {author} {\bibfnamefont {Kenji}\ \bibnamefont {Watanabe}}, \bibinfo {author} {\bibfnamefont {Takashi}\ \bibnamefont {Taniguchi}}, \bibinfo {author} {\bibfnamefont {Efthimios}\ \bibnamefont {Kaxiras}}, \ and\ \bibinfo {author} {\bibfnamefont {Pablo}\ \bibnamefont {Jarillo-Herrero}},\ }\bibfield  {title} {\enquote {\bibinfo {title} {Unconventional superconductivity in magic-angle graphene superlattices},}\ }\href {\doibase 10.1038/nature26160} {\bibfield  {journal} {\bibinfo  {journal} {Nature}\ }\textbf {\bibinfo {volume} {556}},\ \bibinfo {pages} {43--50} (\bibinfo {year} {2018})}\BibitemShut {NoStop}%
\bibitem [{\citenamefont {Di~Liberto}\ \emph {et~al.}(2019)\citenamefont {Di~Liberto}, \citenamefont {Mukherjee},\ and\ \citenamefont {Goldman}}]{di2019nonlinear}%
  \BibitemOpen
  \bibfield  {author} {\bibinfo {author} {\bibfnamefont {Marco}\ \bibnamefont {Di~Liberto}}, \bibinfo {author} {\bibfnamefont {Sebabrata}\ \bibnamefont {Mukherjee}}, \ and\ \bibinfo {author} {\bibfnamefont {Nathan}\ \bibnamefont {Goldman}},\ }\bibfield  {title} {\enquote {\bibinfo {title} {Nonlinear dynamics of {Aharonov-Bohm} cages},}\ }\href {\doibase 10.1103/PhysRevA.100.043829} {\bibfield  {journal} {\bibinfo  {journal} {Phys. Rev. A}\ }\textbf {\bibinfo {volume} {100}},\ \bibinfo {pages} {043829} (\bibinfo {year} {2019})}\BibitemShut {NoStop}%
\bibitem [{\citenamefont {Gligori{\'c}}\ \emph {et~al.}(2019)\citenamefont {Gligori{\'c}}, \citenamefont {Beli{\v{c}}ev}, \citenamefont {Leykam},\ and\ \citenamefont {Maluckov}}]{gligoric2019nonlinear}%
  \BibitemOpen
  \bibfield  {author} {\bibinfo {author} {\bibfnamefont {Goran}\ \bibnamefont {Gligori{\'c}}}, \bibinfo {author} {\bibfnamefont {Petra~P}\ \bibnamefont {Beli{\v{c}}ev}}, \bibinfo {author} {\bibfnamefont {Daniel}\ \bibnamefont {Leykam}}, \ and\ \bibinfo {author} {\bibfnamefont {Aleksandra}\ \bibnamefont {Maluckov}},\ }\bibfield  {title} {\enquote {\bibinfo {title} {Nonlinear symmetry breaking of {Aharonov-Bohm} cages},}\ }\href {\doibase 10.1103/PhysRevA.99.013826} {\bibfield  {journal} {\bibinfo  {journal} {Phys. Rev. A}\ }\textbf {\bibinfo {volume} {99}},\ \bibinfo {pages} {013826} (\bibinfo {year} {2019})}\BibitemShut {NoStop}%
\bibitem [{\citenamefont {Goblot}\ \emph {et~al.}(2019)\citenamefont {Goblot}, \citenamefont {Rauer}, \citenamefont {Vicentini}, \citenamefont {Le~Boit{\'e}}, \citenamefont {Galopin}, \citenamefont {Lema{\^\i}tre}, \citenamefont {Le~Gratiet}, \citenamefont {Harouri}, \citenamefont {Sagnes}, \citenamefont {Ravets} \emph {et~al.}}]{goblot2019nonlinear}%
  \BibitemOpen
  \bibfield  {author} {\bibinfo {author} {\bibfnamefont {V}~\bibnamefont {Goblot}}, \bibinfo {author} {\bibfnamefont {B}~\bibnamefont {Rauer}}, \bibinfo {author} {\bibfnamefont {F}~\bibnamefont {Vicentini}}, \bibinfo {author} {\bibfnamefont {A}~\bibnamefont {Le~Boit{\'e}}}, \bibinfo {author} {\bibfnamefont {Elisabeth}\ \bibnamefont {Galopin}}, \bibinfo {author} {\bibfnamefont {A}~\bibnamefont {Lema{\^\i}tre}}, \bibinfo {author} {\bibfnamefont {L}~\bibnamefont {Le~Gratiet}}, \bibinfo {author} {\bibfnamefont {A}~\bibnamefont {Harouri}}, \bibinfo {author} {\bibfnamefont {I}~\bibnamefont {Sagnes}}, \bibinfo {author} {\bibfnamefont {S}~\bibnamefont {Ravets}},  \emph {et~al.},\ }\bibfield  {title} {\enquote {\bibinfo {title} {Nonlinear polariton fluids in a flatband reveal discrete gap solitons},}\ }\href {\doibase 10.1103/PhysRevLett.123.113901} {\bibfield  {journal} {\bibinfo  {journal} {Phys. Rev. Lett.}\ }\textbf {\bibinfo {volume} {123}},\ \bibinfo {pages} {113901} (\bibinfo {year} {2019})}\BibitemShut
  {NoStop}%
\bibitem [{\citenamefont {Pelegr{\'\i}}\ \emph {et~al.}(2020)\citenamefont {Pelegr{\'\i}}, \citenamefont {Marques}, \citenamefont {Ahufinger}, \citenamefont {Mompart},\ and\ \citenamefont {Dias}}]{pelegri2020interaction}%
  \BibitemOpen
  \bibfield  {author} {\bibinfo {author} {\bibfnamefont {G}~\bibnamefont {Pelegr{\'\i}}}, \bibinfo {author} {\bibfnamefont {AM}~\bibnamefont {Marques}}, \bibinfo {author} {\bibfnamefont {V}~\bibnamefont {Ahufinger}}, \bibinfo {author} {\bibfnamefont {J}~\bibnamefont {Mompart}}, \ and\ \bibinfo {author} {\bibfnamefont {RG}~\bibnamefont {Dias}},\ }\bibfield  {title} {\enquote {\bibinfo {title} {Interaction-induced topological properties of two bosons in flat-band systems},}\ }\href {\doibase 10.1103/PhysRevResearch.2.033267} {\bibfield  {journal} {\bibinfo  {journal} {Phys. Rev. Res.}\ }\textbf {\bibinfo {volume} {2}},\ \bibinfo {pages} {033267} (\bibinfo {year} {2020})}\BibitemShut {NoStop}%
\end{thebibliography}

%

\end{document}